\documentclass{cargese}\usepackage{graphicx}
\let\footnote\savefootnote
\let\footnotetext\savefootnotetext 
\let\footnoterule\savefootnoterule 
\normallatexbib
\def\mathbb#1{ {\bf #1} }

\begin{document}

%------------ article title  ------------------->>
% For a long title use \\ to cut lines.
% In that case, supply  alternate version of the title
% in square brackets, (it will go in the Table of contents during final
% production of the book.
% \articletitle[Short title]{The long version \\ of this title}

\articletitle[Loewner Chains.]{Loewner Chains.}

%% optional, to supply a shorter version of the title for the running head:
%%\chaptitlerunninghead{}

\author{Michel BAUER and Denis BERNARD}

%% multiple authors at the same institution may be separated with \\
%% like in \author{Samuel Bostaph\\
%%                 and Gregor Kariotis}

%% Your Institution and address. May cut into  separated lines with \\

\affil{Service de Physique Th\'eorique de Saclay  
\footnote{CEA/DSM/SPhT, Unit\'e de recherche associ\'ee au CNRS, URA 2306 du CNRS},\\
CEA-Saclay, 91191 Gif-sur-Yvette, France}

% optional email address
\email{michel.bauer@cea.fr, dbernard@spht.saclay.cea.fr}

%% Repeat the above for multiple authors at different institutions.
%% \author{ }
%% \affil{ }
%% \email{ }

% optional abstract
\begin{abstract}
  These lecture notes on 2D growth processes are divided in two parts.
  The first part is a non-technical introduction to stochastic Loewner
  evolutions (SLEs). Their relationship with 2D critical interfaces is
  illustrated using numerical simulations.  Schramm's argument mapping
  conformally invariant interfaces to SLEs is explained. The second
  part is a more detailed introduction to the mathematically
  challenging problems of 2D growth processes such as Laplacian
  growth, diffusion limited aggregation (DLA), etc. Their description
  in terms of dynamical conformal maps, with discrete or continuous
  time evolution, is recalled.  We end with a conjecture based on
  possible dendritic anomalies which, if true, would imply that the
  Hele-Shaw problem and DLA are in different universality classes.
\end{abstract}

%------------ body of article ------------------->>
% Write your article here. 
% Note that the \section{section title}
% command allows for the form \section[short title]{very long\\ title}
% Idem for \subsection and \subsubsection
%------------ end of article ------------------->>

%% \begin{figure}
%% \begin{center}
%% \includegraphics[scale=.7]{open.eps}
%% \label{open}
%% \caption{A random picture.}
%% \end{center}
%% \end{figure}

Growth phenomena are ubiquitous in the physical world at many scales,
from crystals to plants to dunes and larger. They can be studied in
many frameworks, deterministic of probabilistic, in discrete or
continuous space and time. Understanding growth is usually a very
difficult task.  This is true even in two dimensions, the case we
concentrate on in these notes.

Yet two dimensions is a highly favorable situation because it allows
to make use of the power of complex analysis in one variable. In many
interesting cases, the growing object in two dimensions can be seen as
a domain, i.e. a contractible open subset of the Riemann sphere (the
complex plane with a point at infinity added). A deep theorem of
Riemann asserts that such a domain, whatever complicated and fancy, is
conformally equivalent to a simple reference domain, which is usually
taken as the upper-half plane or the unit disk. This conformal
equivalence is unique once an appropriate normalization, which may
depend on the growth problem at hand, has been chosen. Cauchy's
theorem allows to write down an integral representation for the
conformal map as an integral along the boundary of the reference
domain, involving a density. This density is time dependent. Then the
time derivative of the conformal map has an analogous representation
and a nice way to specify the growth rule is often directly on this
density. This leads to the concept of Loewner chains, which is the
central theme of these notes. We shall illustrate Loewner chains in
several situations.

Our aim is to give a pedagogical introduction to a beautiful subject.
We wanted to show that it leads to many basic mathematical structures
whose appearance in the growth context is not so easy to foresee, like
Brownian motion, integrable systems and anomalies to mention just a
few. We have also tried to stress that some growth processes have
rules which are easy to simulate on the computer. A few minutes of CPU
are enough to get an idea of the shape of the growing patterns, to be
convinced that something interesting and non trivial is going on, and
even sometimes to get an idea of fractal dimensions. This is of course
not to be compared with serious large scale simulations, but it is a
good illustration of the big contrast between simple rules, complex
patterns and involved mathematical structures. However, other growth
models, and among those some have been conjectured to be equivalent to
simple ones, have resisted until recently to precise numerical
calculations due to instabilities.

To avoid any confusion, let us stress that being able to describe a
growth process using tools from complex analysis and conformal geometry
does not mean that the growth process itself is conformally invariant
at all. Conformal invariance of the growth process itself puts rather
drastic conditions on the density that appears in the Loewner chain.

This is illustrated by the first part of these notes, which deals with
conformally invariant interfaces and their relation to stochastic
Loewner evolutions. This part is an elaboration of the the main points
developed during the lectures. 

The second part is an introduction to a larger class of processes
describing the growth of possibly random fractal planar domains and a
review of some of their basic properties. Due to lack of time, this
part was not presented during the lectures.

\vspace{.5cm}

The study of the continuum limit of non-intersecting curves on the
lattice has been a subject of lasting interest both in mathematics and
in physics. A famous example is given by self-avoiding random walks. The
motivation comes from combinatorics, but also from statistical
mechanics. Two dimensions is most interesting because a
non-intersecting curve is the boundary between two domains and can
very often be interpreted as an interface separating two coexisting
phases.

At a critical point and for short range interactions, such interfaces
are expected to be conformally invariant. The argument for that was
given two decades ago in the seminal paper on conformal field theory
\cite{bpz}. The rough idea is the following. At a critical point, a
system becomes scale invariant.  If the interactions on the lattice
are short range, the model is described in the continuum limit by a
local field theory and scale invariance implies that the stress
tensor is traceless. In two dimensions this is enough to
ensure that the theory transforms simply --no dynamics is involved,
only pure kinematics-- when the domain where it is defined is changed
by a conformal transformation. 

The local fields are classified by representations of the infinite
dimensional Virasoro algebra and this dictates the way correlation
functions transform. This has led to a tremendous accumulation of
exact results using conformal field theory (CFT). A situation that is
well under control is that of unitary minimal models. The Hilbert
space of the system splits as a finite sum of representations of the
Virasoro algebra, each associated to a (local) primary field, and the
corresponding correlation functions can be described rather
explicitly. The study of non-local objects like interfaces at
criticality has not seen such a systematic development and only
isolated though highly artful results \cite{cardy,watts,duplantier}
have been discovered using conformal field theory techniques.

Non local objects like interfaces are not classified by
representations of the Virasoro algebra but the reasoning that led O.
Schramm to the crucial breakthrough \cite{schramm}, i.e. the
definition of stochastic Loewner evolutions (SLEs), rests on a fairly
obvious but cleverly exploited statement of what conformal invariance
means for an interface. Surprisingly it allows to turn this problem
into growth problem, something which looks natural only a posteriori.
For some years, probabilistic techniques have been applied to
interfaces, leading to a systematic understanding that was lacking on
the CFT side. A sample, surely biased by our ignorance, can be found
in refs.\cite{lsw,rohdeschramm,beffara,dubedat,course}. There is now a
satisfactory understanding of interfaces in the continuum limit.
However, from a mathematical viewpoint, giving proofs that a discrete
interface on the lattice has a conformally invariant limit remains a
hard challenge and only a handful of cases has been settled up to now.

There is now a good explanation of the --initially mysterious--
relationship between SLE and CFT. This was one of the main topics of
the lectures. Though this aspect is mostly due to the work of the
present authors (\cite{bibi}, but see also \cite{wernerfrie}), we have
decided to be very sketchy. The interested reader is referred to the
literature.

\vspace{.5cm}

\begin{figure}[htbp]
\begin{center}
\includegraphics[width=0.5\textwidth]{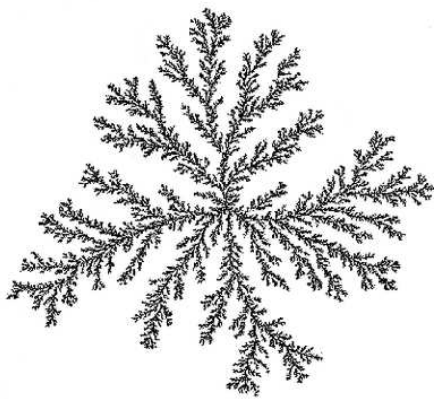}
\caption{An example of DLA cluster obtained by iterating conformal maps.}
     \label{fig:dlaconf}
 \end{center}
\end{figure}

Stochastic Loewner evolution is a simple but particularly interesting
example of growth process for which the growth is local and continuous
so that the resulting set is a curve without branching.  Of course
other examples have been studied in connection with 2d physical
systems. The motivations are sometimes very practical. For instance,
is it efficient to put a pump in the center of oil film at the surface
of the ocean to fight against pollution? The answer has to do with the
Laplacian growth or Hele-Shaw problem \cite{heleshaw}. The names
diffusion limited aggregation \cite{dla}, see Figure (\ref{fig:dlaconf}),
and dielectric breakdown \cite{dielec} speak for themselves. Various
models have been invented, sometimes with less physical motivation,
but in order to find more manageable growth processes. These include
various models of iterated conformal maps \cite{lgdiscret}, etc. See
refs.\cite{revue} for recent reviews.  As mentioned before, in most
cases the shape of the growing domains is encoded in a uniformizing
conformal map whose evolution describes the evolution of the domain.
The dynamics can be either discrete or continuous in time, it can be
either deterministic or stochastic. But the growth process is always
described by a simple generalization of the Loewner equation called a
Loewner chain.

The general understanding of these Loewner chains is still embryonic,
in strong contrast with the case of Loewner evolutions, which deal
with local growth. For many problems the most basic question are
unanswered. For instance, as we shall explain below, Laplacian growth
is at the same time a completely integrable system and an ill-posed
problem because it develops singularities in finite time.
Hydrodynamics gives a natural regularization via the introduction of
surface tension. It has been argued for some time that in the limit of
vanishing surface tension one retrieves a model which is in the same
universality class as diffusion limited aggregation. But the
experimental and numerical evidence is inconclusive and there is no
consensus. In fact, at the end of these notes we shall give an
argument suggesting that they belong to different universality
classes. This is very conjectural, but as should be clear from this
introduction, non-local growth processes are certainly a source of
interesting and challenging problems.

Table of contents:\\
-- Section 1: Critical interfaces and stochastic processes.\\
-- Section 2: Loewner chains.\\

%\newpage

\section{Critical interfaces and stochastic processes}

The following pedagogical introduction to conformally invariant random
curves breaks in two part. The first is a list of examples of critical
interfaces on the lattice and some of their properties. The second is
a presentation of O. Schramm's derivation of SLE.

In this part the upper-half plane is used as a reference geometry. 

\subsection{Three examples and a generality}

Let us start with three examples. Our aim is to explain their
definitions, to show a few samples and, in the first two cases, to give a
numerical estimate of the fractal dimension.

The first, loop-erased random
walks, belongs to the realm of ``pure combinatorics''. 

The second, percolation, is at the frontier between pure combinatorics
and statistical mechanics because it arises very naturally as the domain
boundary of a statistical mechanics model, but the Boltzmann weights are
trivial. It is however a limiting case of a family of models with non
trivial Boltzmann weights and, from that point of view, quantities
relevant for percolation are obtained by exploring an infinitesimal
neighborhood around the trivial Boltzmann weights. 

The third example, the Ising
model, is deeply rooted in statistical mechanics. 

We shall end this section by abstracting a crucial property of
interfaces which allows to make an efficient use of conformal
invariance and derive the Markov property for SLEs.

\subsubsection{Loop-erased random walks}

This example is purely of combinatorial nature. A loop-erased random
walk is a random walk with loops erased along as they appear. More
formally, if $X_0,X_1,\cdots,X_n$ is a finite sequence of abstract
objects, we define the associated loop-erased sequence by the
following recursive algorithm.

\vspace{.2cm}

\noindent Until all terms in the sequence are distinct, 

{\bf Step 1} Find the couple $(l,m)$ with
$0\leq l <m$ such that the terms with indexes from $0$ to $m-1
$ are all distinct but the terms with indexes $m$ and $l$ coincide. 

{\bf Step 2} Remove the terms with indexes from $l+1$ to $m$, and shift
the indexes larger than $m$ by $l-m$ to get a new sequence.

\vspace{.2cm}

Let us look at two examples. \\
For the ``month'' sequence $j,f,m,a,m,j,j,a,s,o,n,d$, the first loop
is $m,a,m$, whose removal leads to $j,f,m,j,j,a,s,o,n,d$, then
$j,f,m,j$, leading to $j,j,a,s,o,n,d$, then $j,j$ leading to
$j,a,s,o,n,d$ where all terms are distinct. \\ 
For the ``reverse month'' sequence $d,n,o,s,a,j,j,m,a,m,f,j$, the
first loop is $j,j$, leading to $d,n,o,s,a,j,m,a,m,f,j$, then
$a,j,m,a$ leading to $d,n,o,s,a,m,f,j$. 

This shows that the procedure
is not ``time-reversal'' invariant. Moreover, terms that are within a
loop can survive: in the second example $m,f$, which stands in the
$j,m,a,m,f,j$ loop, survives because the first $j$ is inside the loop
$a,j,m,a$ which is removed first. 

\begin{figure}[htbp]
\begin{center}
\includegraphics[width=0.6\textwidth]{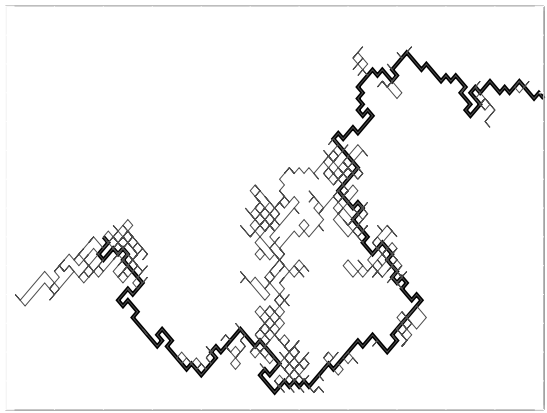} 
\end{center}
\caption{A loop-erased random walk with its shadow.}
     \label{fig:shortLERW-BW}
\end{figure}

A loop-erased random walk is when this procedure is applied to a (two
dimensional for our main interest) random walk. This is very easy to
simulate. Fig.\ref{fig:shortLERW-BW} represents a loop-erased walk of
200 steps. The thin lines build the shadow of the random walk (where
shadow means that we do not keep track of the order and multiplicity
of the visits) and the thick line is the corresponding loop-erased
walk. The time asymmetry is clearly visible and allows to assert with
little uncertainty that the walk starts on the left.

To fit with the general SLE framework, let us restrict to loop-erased
random walks in the upper-half plane. There are a few options for the
choice of
boundary conditions. \\
A first choice is to consider reflecting boundary conditions on the
real axis for the
random walk. \\
Another choice is annihilating boundary conditions: if the random walk
hits the real axis, one forgets everything and starts anew at the
origin. Why this is a natural boundary condition has to wait until
section \ref{subsubsec:generality}.

\begin{figure}[htbp]
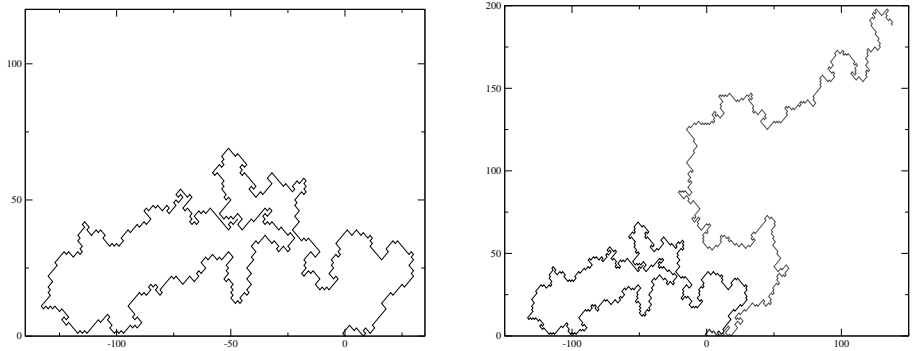

\begin{center}
\includegraphics[width=0.465\textwidth]{bernard.fig7.eps} \hfill
\includegraphics[width=0.47\textwidth]{bernard.fig8.eps}
\end{center}
\caption{On the left: a large loop is about to be created. 
On the right: the massive rearrangement to go from 633 to 634
steps.}
      \label{fig:LERW633-634}
\end{figure}

Due to the fact that on a two-dimensional lattice a random walk is
recurrent (with probability one it visits any site infinitely many
times), massive rearrangement occur with probability one. This means
that if one looks at the loop-erased random walk associated to a given
random walk, it does not have a limit in any sense when the size of
the random walk goes to infinity. Let us illustrate this point. The
samples in fig.\ref{fig:LERW633-634} were obtained with
reflecting boundary conditions. It takes 12697 random walk steps to
build a loop-erased walk of length 633, but step 12698 of the random
walk closes a long loop, and then the first occurrence of a
loop-erased walk of length 634 is after 34066 random walk steps.
Observe that in the mean time most of the initial steps of the loop-erased
walk have been reorganized.

However, simulations are possible because when the length of the
random walk tends to infinity, so does the maximal length of the
corresponding loop-erased walk with probability one.

\vspace{.3cm}

Though annihilating boundary conditions lead to remove even more parts
of the random walk than the reflecting ones, the corresponding process
can be arranged (conditioned in probabilistic jargon) to solve the
problem of convergence as follows.

Instead of stopping the process when the loop-erased walk has reached
a given length, one can stop it when it reaches a certain altitude,
say $n$, along the $y$-axis. Whatever the corresponding random walk
has been, the only thing that matters is the last part of it,
connecting the origin to altitude $n$ without returning to altitude
$0$. Moreover, the first time the loop-erased walk reaches altitude
$n$ is exactly the first time the random walk reaches altitude $n$.
Now a small miracle happens: if a 1d symmetric random walk is
conditioned to reach altitude $n$ before it hits the origin again, the
resulting walk still has the Markov property. It is a discrete
equivalent to the 3d Bessel process (a Bessel process describes the
norm of a Brownian motion, however no knowledge of Bessel processes is
needed here, we just borrow the name). When at site m, $0 < m <n$, the
probability to go to $m\pm1$ is $(1\pm 1/m)/2$, independently of all
previous steps. Observe that there is no $n$ dependence so that we can
forget about $n$, i.e. let it go to infinity.  The discrete 3d Bessel
process is not recurrent and tends to infinity with probability one:
for any altitude $l$ there is with probability one a time after which
the discrete 3d Bessel process remains above $l$ for ever. Henceforth,
we choose to simulate a symmetric simple random walk along the $x$
axis and the discrete 3d Bessel process along the $y$-axis and we
erase the loops of this new process. This leads to the convergence of
the loop-erased walk and numerically to a more economical simulation.

\begin{figure}[htbp]
\begin{center}
\includegraphics[width=0.8\textwidth]{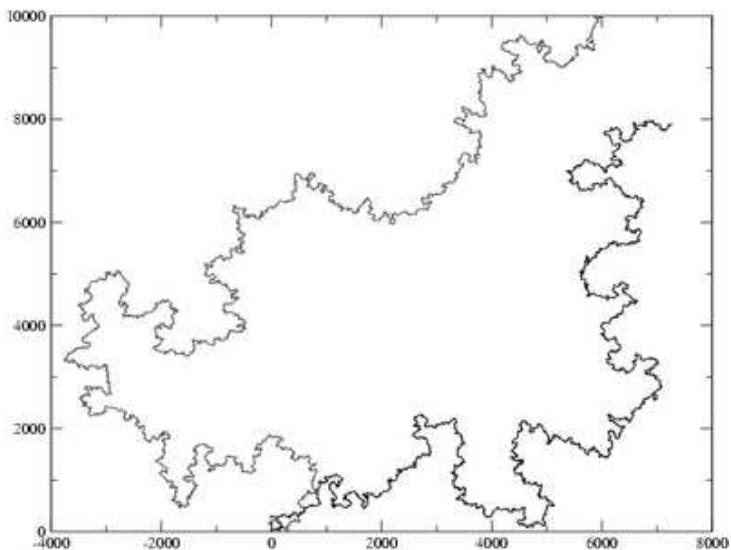}
\end{center}
\caption{A sample of the loop-erased random walk for the two boundary
  conditions.} 
      \label{fig:samplesLERW-BW}
\end{figure}

Fig.\ref{fig:samplesLERW-BW} is a simulation of about $10^5$ steps,
both for reflecting and annihilating boundary
conditions. At first glance, one observes in both cases similar simple
(no multiple points) but irregular curves with possibly fractal
behavior.

To estimate the Hausdorff dimensions in both cases, we have
generated samples of random walks, erased the loops and made the
statistics of the number of steps $S$ of the resulting walks compared
to a typical length $L$ (end-to-end distance for reflecting boundary
conditions, maximal altitude for annihilating boundary conditions). In
both cases, one observes that $S \propto L^{\delta}$ and a modest
numerical effort (a few hours of CPU) leads to $\delta=1.25\pm.01$.
This is an indication that the boundary conditions do not change the
universality class. 

\begin{figure}[htbp]
\begin{center}
\includegraphics[width=0.5\textwidth]{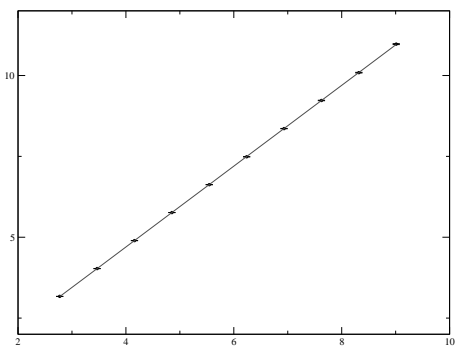}
\end{center}
\caption{The logarithm of the average length of the loop-erased random walk
  versus logarithm of the maximum altitude. The numerical results are
  the circles, the straight line is the linear regression, the error
  bars are shown.}
      \label{fig:moyenne-BW}
\end{figure}

To get an idea of how small the finite size corrections are, observe
fig.\ref{fig:moyenne-BW}. The altitude was sampled from $2^4$ to
$2^{13}$. The best fit gives a slope $1.2496$
and the first two points already give $1.2403$.

As recalled in the introduction, it is believed on the basis of
intuitive arguments that in two dimensions scale invariance implies
conformal invariance, providing there are no long range interactions.
What does this absence of long range interactions mean for loop-erased
random walks?  Clearly along the loop-erased walk there are long range
correlations, if only because a loop-erased random walk cannot cross
itself. However, the relevant feature for the intuitive argument is
that, in the underlying 2d physical space, interactions are indeed
short range. At each time step, the increment of the underlying random
walk is independent of the rest of the walk, and the formation of a
loop to be removed is known from data at the present position of the
random walk.

\vspace{.3cm}

>From the analytical viewpoint, the loop-erased random walk is one of
the few systems that has been proved to have a conformally invariant
distribution in the continuum limit, the fractal dimension being
exactly $5/4$. A naive idea to get directly a continuum limit
representation of loop-erased walks would be to remove the loops from
a Brownian motion.  This turns out to be impossible due to the
proliferation of overlapping loops of small scale. However, the
SLE$_2$ process, to be defined later, gives a direct definition.  In
fact, it is the consideration of loop-erased random walks that led
Schramm \cite{schramm} to propose SLE as a description of interfaces.
 
\subsubsection{Percolation}

\begin{figure}[htbp]
\begin{center}
\includegraphics[width=0.7\textwidth]{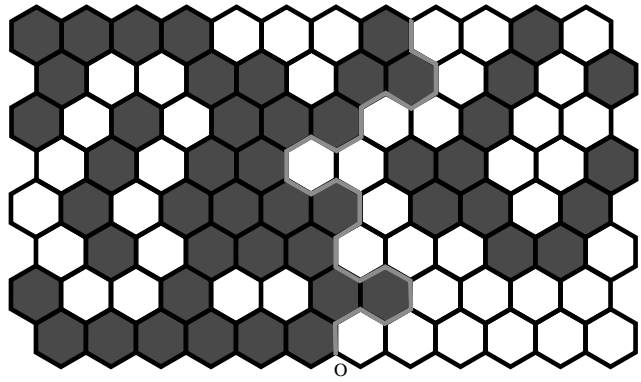}
\caption{The definition of the percolation interface.}
     \label{fig:percodef}
 \end{center}
\end{figure}

To define a random interface for percolation, one possibility is to
pave the upper-half plane with regular hexagons. Each one is,
independently of the others, occupied or empty with probability $1/2$,
with the exception of the ones along the real axis, which are all
empty on the positive real axis and occupied on the negative real
axis, see fig.\ref{fig:percodef}. Then a continuous interface,
starting at the origin and separating occupied and unoccupied
hexagons, is uniquely defined. As before, this defines a simple curve
on the lattice. 

The percolation interface has an obvious but very remarkable property
that singles it out: \textit{locality}. Locality means that the
percolation interface does not depend on the distribution of occupied
and empty sites away from itself. Equivalently, if $\mathbb{D}$ is a
domain (i.e. in this discrete setting a connected and simply connected
family of hexagons in the upper half plane) containing the origin, the
law of the percolation interface in $\mathbb{D}$ before it hits the
boundary of $\mathbb{D}$ for the first time is independent of the
distribution of occupied/empty sites outside $\mathbb{D}$. It is the
same law as for what one would define as the percolation interface in
$\mathbb{D}$ without ever mentioning the world outside $\mathbb{D}$.

\begin{figure}[htbp]
\begin{center}
\includegraphics[width=\textwidth]{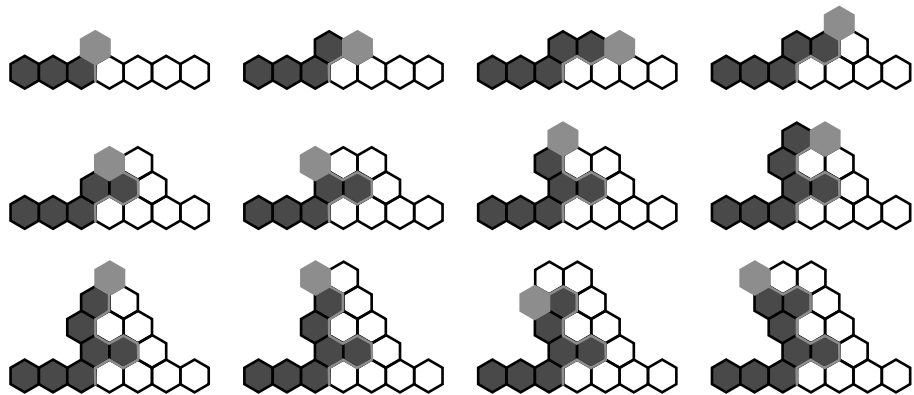}
\caption{The percolation interface as a growth process.}
     \label{fig:percolgrowth}
 \end{center}
\end{figure}

This observation makes the percolation interface easy to simulate. Indeed,
the construction of a percolation interface proceeds inductively, as shown
on the fig.\ref{fig:percolgrowth}.

\begin{figure}[htbp]
\begin{center}
\includegraphics[width=0.85\textwidth]{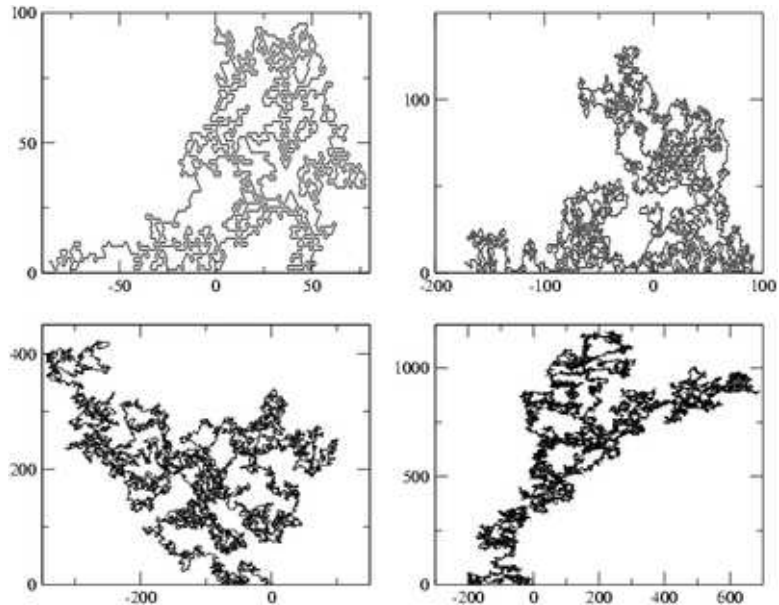}
\caption{Samples of the percolation interface for increasing
  sizes.}
      \label{fig:percosamples}
\end{center}\end{figure}

Fig.\ref{fig:percosamples} shows a few samples of increasing size.

The contrast with the previous example is rather striking. Even for
small samples, the percolation interface makes many twists and turns.
With the resolution of the figure, the percolation interface for large
samples does not look like a simple curve at all! The intuitive
explanation why this does not occur for loop-erased walks is that if
it comes close to itself, then with large probability a few more steps
of the random walk will close a loop.

Getting an estimate of the fractal dimension of the percolation
interface proceeds along lines similar to the first example.  If the
number of steps $S$ of the percolation interface is compared to a
typical length $L$ one observes that $S \propto L^{\delta}$ and a
modest numerical effort (a few hours of CPU) leads to
$\delta=1.75\pm.01$.

The percolation interface is also build by applying rules involving
only a few nearby sites, and again it is expected on general non
rigorous arguments that its scale invariance should imply its
conformal invariance in the continuum limit.

The percolation process is another one among the few systems that has
been proved to have a conformally invariant distribution in the
continuum limit, the fractal dimension being exactly $7/4$. As
suggested by numerical simulations, the continuum limit does not
describe simple curves but curves with a dense set of double points,
and in fact the --to be defined later-- SLE$_6$ process describes not
only the percolation interface but also the percolation hull
\cite{cardy}, which is the complement of the set of points that can be
joined to infinity by a continuous path that does not intersect the
percolation interface.

\subsubsection{The Ising model}

Our next example makes also use of the same pictorial representation:
it is the Ising model on the triangular lattice in the low
temperature expansion. The spins are fixed to be $up$ on the left and
down on the right of the origin along the real axis. The energy of a
configuration is proportional to the length of the curves separating
up and down islands. The proportionality constant has to be
adjusted carefully to lead to a critical system with long range
correlations. This time, making accurate simulations is much more
demanding. On the square lattice, the definition of the interface
suffers from ambiguities, but these become less relevant for larger
sample sizes.  

\begin{figure}[htbp]
\begin{center}
\includegraphics[width=\textwidth]{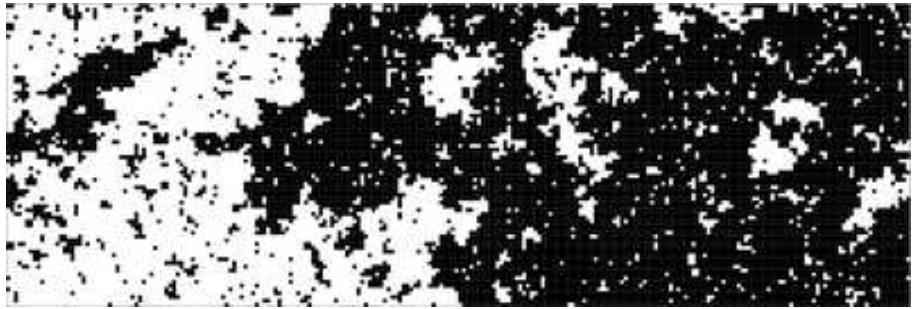}
\caption{A sample for the critical Ising model. The bottom line, where
  the spins are frozen --black on the right, white on the left-- is
  not represented. Courtesy of J. Houdayer.}
     \label{fig:isingsample}
 \end{center}
\end{figure}

Although there is no question that the fractal dimension of the Ising
interface with the above boundary conditions is $11/8$ and is
described by --to be defined later-- SLE$_3$, a mathematical proof
that a continuum limit distribution for the interface exists and is
conformally invariant is still out of reach.

\vspace{.5cm}

We conclude this list of examples with two remarks. 

First, the pictorial representation by a gas of non-intersecting
curves on the hexagonal lattice that we used in the second and the
third example applies to another family of models, the $O(n)$ models,
to which similar interface considerations would apply.

Second, non branching interfaces, described on the lattice by simple
curves, are not the generic situation. For instance the $Q=3$ states
Potts model, three different phases coexist and the physical
interfaces have branch points.

\subsubsection{A generality} \label{subsubsec:generality}

Up to now, we have discussed interfaces in --lattice approximations
of-- the upper half plane. Let us note that they make sense
for more general domains.

\begin{figure}[htbp]
\begin{center}
\includegraphics[width=0.4\textwidth]{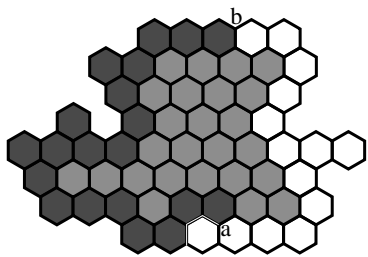}
\end{center}
\caption{A domain on which percolation and the Ising model can be
  defined.}
      \label{fig:percol4}
\end{figure}

Start with the case of the percolation or of the Ising model.  In the
plane, take any connected and simply connected collection of hexagons,
split it into an interior and a connected boundary, and split the
boundary in two connected pieces in such a way that exactly two pairs
of adjacent boundary hexagons, marked $a$ and $b$, carry different
colors, see fig.\ref{fig:percol4}. Observe that we allow a cut which
would correspond to the beginning of the interface.  Hexagons
separated by the cut are not to be counted as neighbors.

Then any coloring of the hexagons in the interior will lead to a
well-defined interface. It will also lead to a well defined energy,
and then the question of the distribution of the interface in such a
domain is meaningful.  For percolation the energy is independent of
the configuration. For the Ising model, it is proportional to the
length of the curves separating up and down sites. Observe that --as
the possible cut separates fixed spins-- counting interactions across
the cut or not in the energy just adds a constant to it, so that it
has no influence on probabilities.

\begin{figure}[htbp]
\begin{center}
\includegraphics[width=0.4\textwidth]{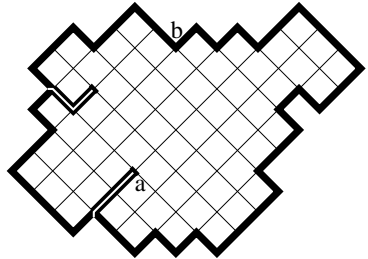}
\end{center}
\caption{A domain on which loop-erased random walks can be defined.}
      \label{fig:lerw1}
\end{figure}

For the loop-erased random case, the idea is similar. One takes a
simply connected piece of the square lattice with two points $a$ and
$b$ on the boundary, again allowing some cuts, see
fig.\ref{fig:lerw1}.
Consider all walks from $a$ to $b$ that do not touch the boundary
except at $a$ before the first step and at $b$ after the last step and
give each such walk of length $l$ a weight $4^{-l}$. Observe that this
choice is exactly the annihilating boundary condition: in the half
plane geometry, $a$ was the origin and $b$ the point at infinity and,
due to the infinite extension of the boundary, we had to go through a
limiting procedure. Then erase the loops to get a probability
distribution for loop-erased random walks from $a$ to $b$ in the
domain. The probability for the simple
symmetric random walk to hit the boundary for the first time at $b$
starting from $a$ can be interpreted as the partition function for
loop-erased walks.

\vspace{.5cm}

We can now go to the point we want to make, valid for all the above
examples.  For any of these, we use $P_{(\mathbb{D},a,b)}$ to denote
the probability distribution for the interface $\gamma_{[ab]}$ from
$a$ to $b$ in $\mathbb{D}$.
  
\begin{figure}[htbp]
\begin{center}
\includegraphics[width=.8\textwidth]{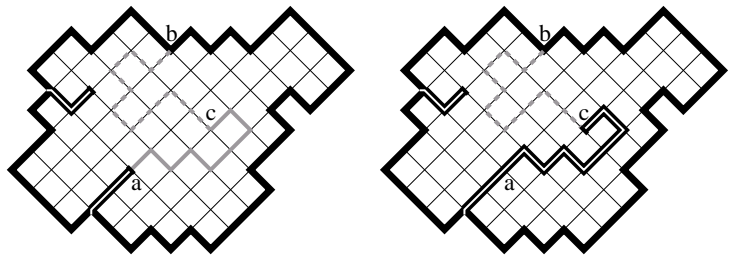}
\end{center}
\caption{An illustration of situations 1) and 2) for the case of
  loop-erased walks. What is the distribution of the dotted curve in
  both situations ?}
      \label{fig:lerw2}
\end{figure}

Suppose that we fix the beginning $\gamma_{[ac]}$ of a
possible interface in domain $\mathbb{D}$, up to a certain point $c$.
Then 1) we can consider the conditional distribution for the rest of
the interface and 2) we can remove the beginning of the interface from
the domain to create a new domain and consider the distribution of the
interface in this new domain. This is illustrated on fig.\ref{fig:lerw2}.

We claim that the distributions defined in 1) and 2) coincide. For
reasons to be explained in a moment, we call this property ``locality at
the interface''. In equations
\[P_{(\mathbb{D},a,b)}(\; .\;|\gamma_{[ac]})=P_{(\mathbb{D}\setminus
  \gamma_{[ac[},c,b)}(\; .\;).\]

It is obvious that these two probabilities are supported on the same
set, namely simple curves along the edges of the lattice, going from
$c$ to $b$ in $\mathbb{D}\setminus \gamma_{[ac[}$. Let us however 
note that for loop-erased random walks, annihilating boundary
conditions are crucial. Reflecting boundary conditions clearly do not
work, if only because the supports do not coincide in that case. 

For the case of loop-erased random walks, the argument for the
equivalence of 1) and 2) goes as follows. Take any random walk
(possibly with loops) $W_0=a,W_1,\cdots,W_l=b$ that contributes to an
interface $\gamma_{[ab]}$ which is $\gamma_{[ac]}$ followed by some
$\gamma_{[cb]}$. Let $m$ be the largest index for which the walk
visits $c$. Because the interface has to start with $\gamma_{[ac]}$,
the walk $W_m=c,\cdots,W_l=b$ cannot cross $\gamma_{[ac[}$ again, so
it is in fact a walk in $\mathbb{D}\setminus \gamma_{[ac[}$ from $c$
to $b$ leading to the interface $\gamma_{[cb]}$. The weight for the
walk $W_0=a,W_1,\cdots,W_l=b$ is $4^{-l}$, i.e. simply the product of
weights for the walks $W_0=a,W_1,\cdots,W_m=c$ and
$W_m=c=a,\cdots,W_l=b$. Then a simple manipulation of weights leads
directly to the announced result.

For the case of percolation and the Ising model, in fact more is true:
we can view $P_{(\mathbb{D},a,b)}$ not only as a probability
distribution for the interface, but as the full probability
distribution for the colors of the hexagons and still check the identity
of 1) and 2). Again, the supports are the same for 1) and 2), namely
any configuration of the colors, except that the colors on both sides
of $\gamma_{[ac]}$ are fixed. For the case of percolation, the colors
are independent of each other so the identity of 1) and 2) is clear.
For the Ising model, the difference is that the conditional
probabilities in 1) take into account the interactions between the
colors along the interface, whereas the probability in 2) does not
take into account the interactions between the spins along the cut
left by the removal of the interface. However, as already mentioned
above, the corresponding colors are fixed anyway, so the Boltzmann
weights for the configurations that are in the support of 1) or 2)
differ by a multiplicative constant, which disappears when
probabilities are computed.

This argument extends immediately to systems with only nearest
neighbor interactions. They can be defined on any graph. If any
subset of edges is chosen and the configuration at both end of each
edge is frozen, it makes no difference for probabilities to consider
the model on a new graph in which the frozen edges have been deleted. 

Instead of looking for further generalizations, we argue more
heuristically that the continuum limit for a system with short range
interactions should satisfy locality at the interface. The use of this
locality property --which, as should be amply evident, has nothing to
do with conformal invariance-- together with the conformal invariance
assumption is at the heart of O. Schramm's derivation of stochastic
Loewner evolutions.

\subsection{O. Schramm's argument}

We break the argument in several pieces. 
The heart of the probabilistic derivation establishes two  properties 
of conformally invariant interfaces:
\begin{itemize}
\item the Markov property, 
\item the stationarity of increments,
\end{itemize}
see sections \ref{subsubsec:markovetstat}, \ref{subsubsec:cii} and
\ref{subsubsec:SLE}. The crucial results needed for the probabilistic
part, in particular the basics of Loewner evolutions, are collected in
\ref{subsubsec:rth} and \ref{subsubsec:LE}. The last two subsections
are not directly related to the argument. They collect some basic
facts on the fractal properties of SLE interfaces and on the
connection with CFT.

\subsubsection{Riemann's theorem and hulls}
\label{subsubsec:rth}
In this section we collect a few indispensable results used in
the rest of the course. 

A domain is a non empty connected and simply connected open set
strictly included in the complex plane $\mathbb{C}$. Simple
connectedness is a notion of purely topological nature which in two
dimensions asserts essentially that a domain has no holes and is
contractible. But it is a deep theorem of Riemann that two domains are
always conformally equivalent, i.e. there is an invertible holomorphic
map between them. For instance, the upper-half plane $\mathbb{H}$ is a
domain. It is well known that it has a three dimensional Lie group of
conformal automorphisms, $PSL_2(\mathbb{R})$, that also acts on the
boundary of $\mathbb{H}$. There is a unique automorphism, possible
followed by a transposition, that maps any triple of boundary points
to any other triple of boundary points. By Riemann's theorem, this is
also true for any other domain, at least if the boundary is not too
wild.

Riemann's theorem is used repeatedly in the rest of this course and is
the starting point of many approaches to growth phenomena in two
dimensions.

For later use, we note that one can be a bit more explicit when the
domain $\mathbb{D}$ differs only locally from the upper half plane
$\mathbb{H}$, that is if $\mathbb{K}= \mathbb{H} \setminus \mathbb{D}$
is bounded. Such a set $\mathbb{K}$ is called a hull. The real points
in the closure of $\mathbb{K}$ in $\mathbb{C}$ form a compact set
which we call $\mathbb{K}_{\mathbb{R}}$.  Let $f: \mathbb{H} \mapsto
\mathbb{D}$ be a conformal bijection. As the boundary of $\mathbb{H}$
is smooth, $f$ has a continuous extension to
$\overline{\mathbb{R}}\equiv \mathbb{R}\cup \infty$, and
$f^{-1}(\overline{\mathbb{R}}\setminus \mathbb{K}_{\mathbb{R}})$ is a
non-empty open set in $\overline{\mathbb{R}}$ with compact complement.
We call the complement the cut of $f$. By the Schwarz symmetry
principle, defining $f(z)=\overline{f(\bar{z}})$ for $\Im{\rm m} \, z
\leq 0$ gives an analytic extension of $f$ to the whole Riemann sphere
minus the cut. Across the cut, $f$ has a purely imaginary nonnegative
discontinuity which we write as a Radon-Nikodym derivative
$d\mu_f/dx$.

One can use the $PSL_2(\mathbb{R})$ automorphism group of
$\mathbb{H}$ to ensure that $f$ is holomorphic at $\infty$ and
$f(w)-w=O(1/w)$ there. This is called the hydrodynamic
normalization. It involves three conditions, so there is no
further freedom left. We shall denote this special representative
by $f_\mathbb{K}$, which is uniquely determined by $\mathbb{K}$:
any property of $f_\mathbb{K}$ is an intrinsic property of $\mathbb{K}$.

Cauchy's theorem yields

\begin{equation} \label{eq:repbydiscont}
f_\mathbb{K}(w)=w+\frac{1}{2\pi}\int_\mathbb{R} 
\frac{d\mu_{f_\mathbb{K}}(x)}{x-w}, 
\end{equation} 

A quantity that plays an important role in the sequel is
$$C_\mathbb{K}\equiv \frac{1}{2\pi}\int_\mathbb{R}
d\mu_{f_\mathbb{K}}(x),$$ a positive (unless $\mathbb{K}=\emptyset$)
number called the capacity of $\mathbb{K}$, which is such that
$f_\mathbb{K}(w)=w-C_\mathbb{K}/w+O(1/w^2)$ at infinity. The
usefulness of capacity stems from its good behavior under
compositions: if $\mathbb{K}$ and $\mathbb{K}'$ are two hulls,
$\mathbb{K}\cup f_\mathbb{K}(\mathbb{K}')$ is a hull and
\begin{equation} \label{eq:addcapa}
C_{\mathbb{K}\cup f_\mathbb{K}(\mathbb{K}')}=C_\mathbb{K}+C_{\mathbb{K}'},
\end{equation}
as seen by straightforward expansion at infinity of $f_\mathbb{K}\circ
f_{\mathbb{K}'}$, the map associated to $\mathbb{K}\cup
f_\mathbb{K}(\mathbb{K}').$ In particular capacity is a continuous
increasing function on hulls.

\vspace{.3cm}

Anticipating a little bit, let us note immediately that giving a
dynamical rule for the evolution of the finite positive measure
$d\mu_{f_\mathbb{K}}(x)$ is a good way to define growth processes.

\subsubsection{Conformally invariant interfaces} 
\label{subsubsec:cii}
Consider a domain $\mathbb{D}$, with two distinct points on its
boundary, which we call $a$ and $b$. A simple curve, denoted by
$\gamma_{[ab]}$, from $a$ to $b$ in $\mathbb{D}$ is the image of a
continuous one-to-one map $\gamma$ from the interval $[0,+\infty]$ to
$\mathbb{D}\cup\{a,b\}$ such that $\gamma(0)=a$, $\gamma(\infty)=b$
and $\gamma_{]ab[}\equiv \gamma(]0,\infty[)\subset \mathbb{D}$.
Alternatively, a simple curve from $a$ to $b$ is an
equivalence class of such maps under increasing reparametrizations. A
point on it has no preferred coordinate but is has a past and a future.
If $c \in \mathbb{D}$ is an interior point, we use a similar
definition for a simple curve $\gamma_{[ac]}$ from $a$ to $c$ in
$\mathbb{D}$.

Note that, apart from the fact that on the lattice we could use
lattice length as a parameter along the interface, we have just
rephrased in the continuum --but with the same notations-- what we did
before in a discrete setting.

\begin{figure}[htbp]
\begin{center}
\includegraphics[width=0.9\textwidth]{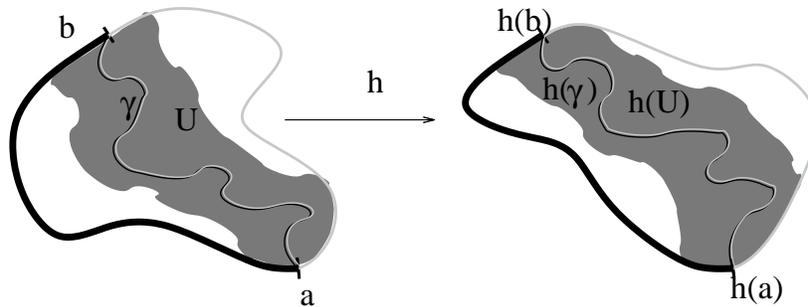}
\caption{Conformal invariance for change of domain.}
     \label{fig:cargchord1}
 \end{center}
\end{figure}

Our aim is to study conformally invariant probability measures on the
set of simple curves from $a$ to $b$ in $\mathbb{D}$.  There is a
purely kinematical step, which demands that if $h$ is any conformal
map that sends $\mathbb{D}$ to another domain $h(\mathbb{D})$, the
measure for $(h(\mathbb{D}),h(a),h(b))$ should be the image by $h$ of
the measure for $(\mathbb{D},a,b)$:
\[P_{(\mathbb{D},a,b)}(\gamma_{[ab]} \subset
U)=P_{(h(\mathbb{D}),h(a),h(b))}(h(\gamma_{[ab]}) \subset h(U)),\]
where $P_{(\mathbb{D},a,b)}(\gamma_{[ab]} \subset U)$ denotes the
  probability for the curve $\gamma_{[ab]}$ to remain in a subset $U$
  of $\mathbb{D}$. See fig.\ref{fig:cargchord1}.

This condition is natural and it is the one that conformal field
theory suggests immediately. Let us note however that a totally
different definition of conformal invariance is understood in the
familiar statement ``two dimensional Brownian motion is conformally
invariant''.

Observe that we could take any measure for $(\mathbb{D},a,b)$ --well,
with the invariance under the one parameter group of automorphisms
that fixes $(\mathbb{D},a,b)$-- and declare that the measure in
$h(\mathbb{D})$ is obtained by definition by the rule above. To make
progress, we need to combine conformal invariance with locality at the
interface.

\subsubsection{Markov property and stationarity of increments}
\label{subsubsec:markovetstat}
This short section establishes the most crucial properties of
 conformally invariant interfaces. 
 
\begin{figure}[htbp]
\begin{center}
\includegraphics[width=0.9\textwidth]{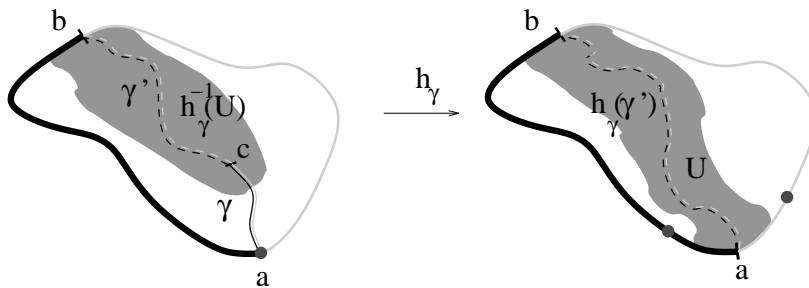}
\caption{Conformal invariance for conditional probabilities.}
     \label{fig:cargchord2}
 \end{center}
\end{figure}

 Take $c\in \mathbb{D}$ and let $\gamma_{[ac]}$ be a simple curve from
 $a$ to $c$ in $\mathbb{D}$. Observe that $\mathbb{D}\setminus
 \gamma_{]ac]}$ is a domain.  To answer the question ``if the
 beginning of the interface is fixed to be $\gamma_{[ac]}$, what is
 the distribution of the rest $\gamma'_{[cb]}$ of the interface ?'' we
 apply locality at the interface to argue that this is exactly the
 distribution of the interface in $\mathbb{D}\setminus \gamma_{]ac]}$.
 We map this domain conformally to $\mathbb{D}$ via a map $h_{\gamma}$
 sending $b$ to $b$ and $c$ to $a$, see fig.\ref{fig:cargchord2}.

Conformal invariance implies that the image measure is the original
measure in $P_{(\mathbb{D},a,b)}$, where $h_{\gamma}(\gamma'_{[cb]})$
is an interface that has forgotten $\gamma_{[ac]}$. To summarize :
\begin{center}
\begin{minipage}[t]{0.9\textwidth}

$h_{\gamma}(\gamma'_{[cb]})$ is independent of $\gamma_{[ac]}$ (the
Markov property) and has the same distribution as the original
interface itself (stationarity of increments).
\end{minipage}
\end{center}

\subsubsection{Local growth and Loewner evolutions}
\label{subsubsec:LE}
The Markov and stationarity of increments property make it plain that to understand
the distribution of the full interface, it is enough to understand the
distribution of a small, or even infinitesimal, initial segment, and
then glue segments via conformal maps.

This calls for a description by differential equations. 

For this purpose, it is very convenient --even if by no means
mandatory-- to have a natural parametrization of interfaces. Using
conformal invariance, we can restrict ourselves to the situation when
$(\mathbb{D},a,b)=(\mathbb{H},0,\infty)$.  If $\gamma_{[0\infty]}$ is
a simple curve from $0$ to $\infty$ in $\mathbb{H}$, and $c$ a point
on it, we know that $\mathbb{H}\setminus \gamma_{]0c]}$ is a domain,
and $\gamma_{]0c]}$ itself is a hull. We use capacity as a
parametrization and define a time parameter by $2t(c)\equiv
C_{\gamma_{]0c]}}$. The factor of $2$ is just historical.  The map $t$
is a continuous increasing function of $c$ varying from $0$ to
$\infty$ in $\bar{\mathbb{R}}$ when $c$ moves from $0$ to $\infty$
along $\gamma_{[0\infty]}$. The inverse map $t \mapsto c(t)$ is
well-defined and gives a parametrization of the curve. So we introduce
an increasing family of hulls $\mathbb{K}_t, t\in [0,\infty[$, by
$\mathbb{K}_t=\gamma_{]0c(t)]}$ which has capacity $2t$. Let $f_t\equiv
f_{\mathbb{K}_t}$ be the conformal homeomorphism from $\mathbb{H}$ to
$\mathbb{H}\setminus \mathbb{K}_t$ normalized to satisfy
$f_t(w)=w-2t/w+O(1/w^2)$ at infinity. Define $g_t: \mathbb{H}\setminus
\mathbb{K}_t \mapsto\mathbb{H} $ to be the inverse of $f_t$. Then
$g_t(z)=z+2t/z+O(1/z^2)$ at infinity.

To study the evolution of the family of hulls $\mathbb{K}_t$, fix $\varepsilon
\geq 0$ and consider the hull $\mathbb{K}_{\varepsilon,t}\equiv
g_t(\mathbb{K}_{t+\varepsilon}\setminus \mathbb{K}_t)$, which has capacity
$2\varepsilon$ by eq.(\ref{eq:addcapa}). Define
$f_{\varepsilon,t}\equiv f_{\mathbb{K}_{\varepsilon,t}}$. Then
$g_t=f_{\varepsilon,t}\circ g_{t+\varepsilon}$ on $\mathbb{H}\setminus
\mathbb{K}_{t+\varepsilon}$. Using the representation of
$f_{\mathbb{K}_{\varepsilon,t}}$ in terms of its discontinuity
eq.(\ref{eq:repbydiscont}), we obtain

\[ 
g_{t+\varepsilon}-f_{\varepsilon,t}\circ
g_{t+\varepsilon}=g_{t+\varepsilon}-g_t=\frac{1}{2\pi}\int_\mathbb{R}
\frac{d\mu_{f_{\varepsilon,t}}(x)}{g_{t+\varepsilon}-x}
\]

We introduce now the notion of local growth which is crucial for
interfaces. When $\varepsilon$ is small, $\mathbb{K}_{\varepsilon,t}$
is a tiny piece of curve and the support of
$d\mu_{f_{\varepsilon,t}}$ is small and becomes a point when
$\varepsilon$ goes to $0$. Measures supported at a point are $\delta$
functions, so there is a point $\xi_t$ such that, as a measure,
$d\mu_{f_{\varepsilon,t}}/dx \sim 2\varepsilon \delta(x-\xi_t)$ as
$\varepsilon \rightarrow 0^+$.  If $\mathbb{K}_t$ is a more general
increasing family of hulls of capacity $2t$, we say that the condition
of local growth is satisfied if the above small $\varepsilon$ behavior
holds.  At first sight, it might seem that local growth is only true
for curves, but this is not true. We shall give an example below.

Letting $\varepsilon \rightarrow 0^+$, from the local growth
condition, we infer the existence of a real function $\xi_t$ such that 
\begin{equation}\label{eq:Loewnerevolution}
\frac{dg_t}{dt}(z)=\frac{2}{g_t(z)-\xi_t}.
\end{equation}
It is useful to look at this equation from a slightly different point
of view, taking the function $\xi_t$ as the primary data. The solutions of
this equation for a given function $\xi_t$ with initial condition
$g_0(z)=z$ is called a Loewner evolution. The image of $\xi_t$ by
$g_t^{-1}$ is the tip of the curve at time $t$.

The cases when the local growth condition is not satisfied are called
Loewner chains, see below. Had we used another parametrization of the
curve, the $2$ in the numerator would be replaced by a positive
function of the parameter along the curve.

Informally, if $\mathbb{K}_t$ is a growing curve, we expect that
$g_{t+\varepsilon}(z)-g_t(z)$ describes an infinitesimal cut. This is
confirmed by the explicit solution of eq.(\ref{eq:Loewnerevolution})
for the trivial case $\xi_t\equiv 0$, which yields $g_t(z)^2=z^2+4t$,
the branch to be chosen being such that at large $z$, $g_t(z) \sim z$.
This describes a growing segment along the imaginary axis. So
intuitively, the simple pole in eq.(\ref{eq:Loewnerevolution})
accounts for the existence of a cut and different functions $\xi_t$
account for the different shapes of curves.

One can also solve the case when $\mathbb{K}_t$ is an arc of circle
starting from the origin. It can be obtained from the trivial solution
above by applying appropriate time dependent $PSL_2(\mathbb{R})$
transformations both to $g_t$ and $z$ and then by a time change to
recover the capacity parametrization. This is an illuminating exercise
that we leave to the reader. Take an arc going from $0$ to $2R$ along
a circle of radius $R$. It turns out that when the arc approaches the
real axis to close a half disk, the function $\xi_t$ has a square root
singularity $\xi_t \propto \sqrt{R^2-2t}$. The capacity remains finite
and goes to $R^2$ and the map itself has a limit
$g_{R^2/2}(z)=z+R^2/(z+R)$ which has swallowed the half disk without
violating the local growth condition. One can start the growth process
again. Making strings of such maps with various values of the radii is
a simple way to construct growing families of hulls that are not
curves and that nevertheless grow locally. Note that a square root
singularity for $\xi_t$ is the marginal behavior: if $\xi_t$ is
H\"older of exponent $ > 1/2$, Loewner evolution yields a simple
curve.

\subsubsection{Stochastic (or Schramm) Loewner evolutions}
\label{subsubsec:SLE}

If we sample locally growing hulls with a certain distribution, we get
an associated random process $\xi_t$. In the case of a conformally
invariant distribution, we have established two crucial properties:
Markov property and stationarity of increments. To finish Schramm's
argument leading to SLE, what remains is to see the implications of
these properties on the distribution of $\xi_t$.

The argument and expressions for the meaning of Markov property and
stationarity of increments involved a map $h$ that mapped the tip of
the piece of interface to the initial marked point $a$ and the
final marked point $b$ to itself. The map $h_t(z)=g_t(z)-\xi_t$ has
the required property when the domain is the upper-half plane with $0$
and $\infty$ as marked points. It behaves like
$h_t(z)=z-\xi_t+2t/z+O(1/z^2)$ at infinity. We infer that for $s >t$,
$h_t(\mathbb{K}_s\setminus \mathbb{K}_t)$ is independent of
$\mathbb{K}_{t'},\; t'\leq t$ (Markov property) and is distributed
like a hull of capacity $s-t=C_{h_t(\mathbb{K}_s\setminus
  \mathbb{K}_t)}$ (stationarity of increments).

The hull determines the corresponding map $h$, so this can be rephrased
as: $h_s\circ h_t^{-1}$ (which uniformizes $h_t(\mathbb{K}_s\setminus
\mathbb{K}_t)$) is independent of $h_{t'},\; t'\leq t,$ and distributed
like an $h_{s-t}$.  As $h_s\circ
h_t^{-1}=z-(\xi_s-\xi_t)+(s-t)/z+O(1/z^2)$ at infinity, the driving
parameter for the process $h_s\circ h_t^{-1}$ is $\xi_s-\xi_t$. To
summarize:

\begin{center}
\begin{minipage}[t]{0.9\textwidth}
the Markov property and stationarity of increments for the
interface lead to the familiar statement for the process $\xi_t$: for
$s >t$, $\xi_s-\xi_t$ is independent of $\xi_{t'},\; t'\leq t,$
(Markov property) and distributed like a $\xi_{s-t}$ (stationarity of
increments).
\end{minipage}
\end{center}

To conclude, a last physical input is needed: one demands that the
interface does not branch, which means that at two nearby times the
growth is at nearby points. This implies that $\xi_t$ is a continuous
process, in the sense that it has continuous trajectories. 

One is now in position to apply a mathematical theorem: a 1d
Markov process with continuous trajectories and stationary increments
is proportional to a Brownian motion. We conclude that there is a real
positive number $\kappa$ such that $\xi_t=\sqrt{\kappa} B_t$ for some
normalized Brownian motion $B_t$ with covariance
$\mathbb{E}[B_sB_t]=\min(s,t)$. The same argument without imposing that the
time parametrization is given by the capacity of the hull would
lead to the conclusion that the driving parameter is a continuous
martingale, which is nothing but a Brownian motion after a possibly
random time change. 

A solution of 
\begin{equation}
\label{eq:slec}
\frac{dg_t}{dt}(z)=\frac{2}{g_t(z)-\sqrt{\kappa} B_t}
\end{equation}
is called a chordal Schramm-Loewner evolution of parameter
$\kappa$, in short a chordal SLE$_{\kappa}$.
The connection of this equation with interfaces relies mainly on
conformal invariance. But local growth, absence of branches and to a
lower level locality at the interface also play a crucial role.

\subsubsection{Miscellanea on SLE}

Up to now we have only discussed the situation when the interface goes
from $0$ to $\infty$ in the upper half plane, or more generally from
one point on the boundary of a domain to another one.  This is called
chordal SLE. The argument can be extended easily to two other
situations. The first is for an interface starting at a fixed point on
the boundary of a domain and ending at a fixed point in the bulk. This
is called radial SLE. For the second situation, three points are
chosen on the boundary and the interface starts from the first point
and ends at a random point between the two others. This is called
dipolar SLE. Just as the upper-half plane is a convenient geometry for
chordal SLE, a semi infinite cylinder is nice for radial SLE and an
infinite strip is nice for dipolar SLE.

For radial SLE on a cylinder of circumference $\pi$, the equation reads
\begin{equation}
\label{eq:sler}
\frac{dg_t}{dt}(z)=\frac{2}{\tan (g_t(z)-\sqrt{\kappa} B_t)}
\end{equation}
and for dipolar SLE on a strip of width $\pi/2$, one has
\begin{equation}
\label{eq:sled}
\frac{dg_t}{dt}(z)=\frac{2}{\tanh (g_t(z)-\sqrt{\kappa} B_t)}
\end{equation}

The generalization of SLE to Riemann surfaces with moduli is easy
for the annulus, but raises non-trivial problems in general,
currently subject of active research.

The set of exact results obtained for SLE forms an impressive body of
knowledge. We shall not mention the ones dealing with the explicit
computation of certain crossing probabilities but we list just a few
``pictorial'' properties with some comments. They (the properties and
the comments) should be understood with the standard proviso ``almost
surely'' or ``with probability $1$''.

We start with a surprisingly difficult result\cite{schramm,lsw,rohdeschramm,beffara}.
\begin{itemize}
\item  Whatever the value of
$\kappa$, the pre-image of the driving parameter $\lim_{w \rightarrow
  \sqrt{\kappa} B_t}g_t^{-1}(w)$ is a continuous curve $\gamma_t$,
called the SLE trace. The trace never crosses itself. This property is
crucial if the trace is to be interpreted as a curve separating two
phases.
\item For $\kappa\in [0,4]$ the SLE trace is a simple curve. For
  $\kappa \in ]4,8[$, it has double points. For $\kappa \in [8,
  \infty[$, it is space filling.
\item The fractal dimension $d_{\kappa}$ of the trace is $1+\kappa /8$
  for $\kappa \leq 8$ and $2$ for $\kappa \geq 8$.
\end {itemize}
Using the formula for the dimension of the trace and confronting with
the numerical simulations, it is plausible (actually, these are among
the few cases for which a mathematical proof exists) that loop-erased random
walks correspond to $\kappa=2, d=5/4$ and percolation to $\kappa=6,
d=7/4$. This is also compatible with the general shape of the
numerical samples, which indicate that loop-erased random walks indeed
lead to simple curves and that percolation doesn't. 
 
The hull $\mathbb{K}_t$ is by definition $\mathbb{H}\setminus
g_t^{-1}(\mathbb{H})$. It has the following properties
\begin{itemize}
\item The hull $\mathbb{K}_t$ is the complement of the connected
  component of $\infty$ in $\mathbb{H}\setminus \gamma_{]0,t]}$.
\item For $\kappa\in [0,4]$, the SLE hull is a simple curve coinciding with
the trace. For $\kappa \in ]4,\infty[$, the SLE hull has an nonempty
connected and relatively dense interior. 
\end{itemize}
This may seem surprising at first sight. It is the sign that for $\kappa >
4$, the drift $\sqrt{\kappa} B_t$ goes fast enough for the swallowing
procedure to take place, as described in the closing arc example, but
on all scales.

\begin{figure}[htbp]
\begin{center}
\includegraphics[width=0.9\textwidth]{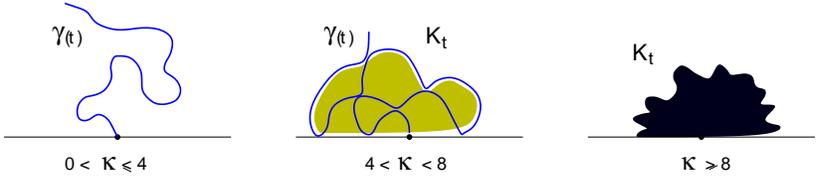}
\caption{ The phases of SLE.}
     \label{fig:sletrace}
 \end{center}
\end{figure}

This is summarized by fig.\ref{fig:sletrace}.

For the many other properties known about SLE, the reader is invited
to read the vast literature
\cite{schramm,lsw,rohdeschramm,beffara,dubedat,course}.   

\subsection{Short remarks on SLE and CFT}

As already mentioned, it is by no means an easy task to prove
that a discrete random curve converges to an SLE. For geometric models
like percolation, loop-erased random walks, self avoiding walks, it is
often easier to compute the appropriate $\kappa$ using known special
properties of the discrete model that are expected to survive in the
continuum limit. 

For instance, one can formulate locality in the
continuum. Showing that SLE$_{\kappa}$ has the locality property
only for $\kappa=6$ is a standard computation in stochastic Ito
calculus, see e.g. \cite{course}.

For critical statistical mechanics models with non trivial
Boltzmann weights, things are more complicated. The system contains
local dynamical degrees of freedom independently from the interface
and, in the continuum limit, these local degrees of freedom are
expected to be described by a conformal field theory (CFT). The most
basic parameter of a CFT is its central charge $c$. Computing it
(even non rigorously) from the lattice model can be a challenge
comparable to the one of computing $\kappa$.  

On the other hand, the relation between $c$ and $\kappa$
can be worked out in general.  The idea is elementary. From a
technical viewpoint, it rests on assumptions similar to the ones needed
in O. Schramm's argument. 

In the discrete setting, take  $O$ to be any observable. If one computes
$\langle O \rangle_{\mathbb{D}\setminus \gamma_{]ac]}}$ in
$\mathbb{D}\setminus \gamma_{]ac]}$, i.e. with part of the interface
fixed, and then averages over $\gamma_{]ac]}$ one retrieves $\langle O
\rangle_{\mathbb{D}}$:
\[ \mathbb{E}[\langle O \rangle_{\mathbb{D}\setminus \gamma_{]ac]}}]
=\langle O \rangle_{\mathbb{D}}\] 
where the expectation is the average
over $\gamma$, see the last reference in \cite{bibi}. This is a
straightforward application of the usual rules of statistical
mechanics

This basic property, which we call the martingale property because in
probabilistic jargon it would be translated as ``$\langle O
\rangle_{\mathbb{D}\setminus \gamma_{]ac]}}$ is a closed martingale
for any observable $O$'' is expected to survive in the continuum
limit. In this limit, one has two powerful tools inherited from
conformal invariance at hand, CFT to compute correlators and SLE to
average over the piece of interface. Doing this for arbitrary values
of $\kappa$ and $c$ means mixing the degrees of freedom from two
different models and there is a priori no reason for the martingale
property to hold.  An explicit computation shows that it holds only if
$$2\kappa c=(6-\kappa)(3\kappa-8).$$
The martingale property also
gives information on the operator content of the CFT. As the discrete
statistical mechanics examples show, there is a change in boundary
conditions at the tip of the interface. The martingale property allows
to identify this boundary changing condition operator as a primary
field of weight $h=(6-\kappa)/(2\kappa)$ degenerate at level two.

This approach also exhibits a family of martingales which are at the
heart of many probabilistic computations. For instance, it gives a
systematic way to interpret probabilities for SLE events as
correlation functions of a CFT, and shows how the changes of behavior
of the trace at $\kappa=4,8$ are related to operator product
expansions. The interested reader is refereed to \cite{bibi}.

%\newpage

\section{Loewner chains}

This section deals with more general 2D growth processes.  Although,
they do not fulfill the local growth and conformal invariance properties of
SLEs, they are nevertheless described by dynamical conformal maps. We
first present systems whose conformal maps have a time continuous
evolution and give examples. We then go on by presenting a discrete
version thereof in terms of iterated conformal maps. We explain
integrability of Laplacian growth. The last part is a discussion
concerning the limit of small ultraviolet cutoff and the consequences
of possible dendritic anomalies.

\subsection{Continuous Loewner chains}

In this part, the exterior of the unit disk is used as the reference
geometry.

\subsubsection{Radial Loewner chains}

\begin{figure}[htbp]
\begin{center}
\includegraphics[width=0.9\textwidth]{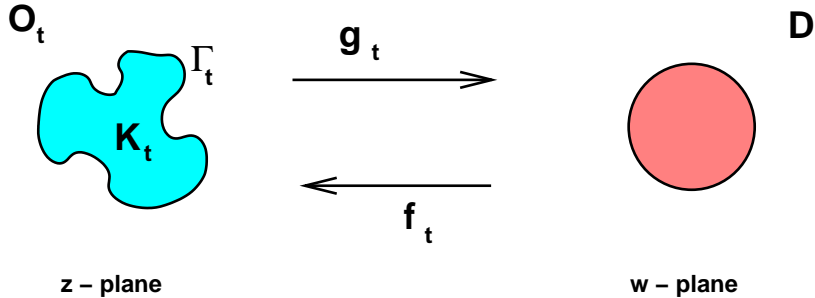}
\caption{Uniformizing maps intertwining the `physical' $z$-plane and
  the `mathematical' $w$-plane.}
     \label{fig:unif}
 \end{center}
\end{figure}

Let $\mathbb{K}_t$ be a family of growing closed planar domains with
the topology of a disk.  Let $\mathbb{O}_t\equiv
\mathbb{C}\setminus\mathbb{K}_t$ be their complements in the complex
plane. See Figure (\ref{fig:unif}).  To fix part of translation
invariance we assume that the origin belongs to $\mathbb{K}_t$ and the
point at infinity to $\mathbb{O}_t$.

Loewner chains describe the evolution of family of conformal maps
$f_t$ uniformizing $\mathbb{D}=\{w\in \mathbb{C};\ |w|>1\}$ onto
$\mathbb{O}_t$. It thus describes the evolution of the physical domains
$\mathbb{O}_t$. We normalize the maps
$f_t:\mathbb{D}\to \mathbb{O}_t$ by demanding that they fix the point at
infinity, $f_t(\infty)=\infty$ and that $f_t'(\infty)>0$.  With $t$
parameterizing time, Loewner equation reads:
\begin{eqnarray}
\frac{\partial}{\partial t} f_t(w) = wf_t'(w)\,
\oint \frac{du}{2i\pi u} \,
\Big({\frac{w+u}{w-u}}\Big)\,\rho_t(u)
\label{loew}
\end{eqnarray}
The integration is over the unit circle 
$\{u\in\mathbb{C}, |u|=1\}$. The Loewner density $\rho_t(u)$ codes
for the time evolution. It may depends on the map $f_t$ in which case
the growth process in non-linear.
For the inverse maps $g_t\equiv f_t^{-1}:\mathbb{O}_t\to \mathbb{D}$, 
Loewner equation reads:
\begin{eqnarray}
\frac{\partial}{\partial t} g_t(z) = -g_t(z)
\oint\frac{du}{2i\pi u} 
\Big({\frac{g_t(z)+u}{g_t(z)-u}}\Big)\,\rho_t(u)
\label{loewbis}
\end{eqnarray}

The behavior of $f_t$ at infinity fixes a scale since at infinity,
$f_t(w)\simeq R_tw+O(1)$ where $R_t>0$, with the dimension of a {\tt
  [length]}, is called the conformal radius of $\mathbb{K}_t$ viewed
from infinity.  $R_t$ may be used to analyze scaling behaviors, since
Kobe 1/4-theorem (see e.g. \cite{conway}) ensures that $R_t$ scales as
the size of the domain.  In particular, the (fractal) dimension $D$ of
the domains $\mathbb{K}_t$ may be estimated by comparing their area
$\mathcal{A}_t$ with their linear size measured by $R_t$:
$\mathcal{A}_t\asymp R_t^{D}$ for large $t$ -- the proportionality
factor contains a cutoff dependence which restores naive dimensional
analysis.

%\vskip 1.0 truecm

\subsubsection{The boundary curves}

The boundary curve $\Gamma_t\equiv \partial \mathbb{O}_t$ is the image of
the unit circle by $f_t$.  We may parameterize the
boundary points by $\gamma_{t;\alpha}=f_t(u)$ with $u=e^{i\alpha}$.
The Loewner equation codes for the evolution of the shape of $\mathbb{O}_t$
and thus for the normal velocity of the boundary points.  Only the
normal velocity is relevant as the tangent velocity is
parameterization dependent.  The tangent to the curve is $\tau = iu
f_t'(u)/|f_t'(u)|$ and the outward normal is $n=-i\tau$ so that the
normal velocity at $\gamma_t$ is $v_n=\Re{\rm e}[\bar n\,
\partial_tf_t(u)]$, or
$$ v_n = |f_t'(u)|\, \Re{\rm e}[\partial_tf_t(u)/ uf_t'(u)].$$

The r.h.s. is determined by the Loewner equation (\ref{loew}) because
this equation may be viewed as providing the solution of a boundary
value problem.  Indeed, recall that for $\hat h(u)$ a real function on
the unit circle, $h(w)=\oint\frac{du}{2i\pi u}
\Big(\frac{w+u}{w-u}\Big)\hat h(u)$ is the unique function analytic
outside the unit disk whose real part on the unit circle is
$\hat h$, i.e.  $\Re{\rm e}[\,h(u)\,]= \hat h(u)$. Thus, since
$\partial_tf_t(w)/wf_t'(w)$ is analytic in $\mathbb{D}$, the Loewner
equation (\ref{loew}) is equivalent to: 
$$ v_n = |f_t'(u)|\, \rho_t(u)$$
or more  explicitly \footnote{For SLE this equation
  has to be modified according to It\^o calculus}:
\begin{eqnarray}
(\partial_\alpha \gamma_{t;\alpha})\,
(\partial_t \overline{\gamma_{t;\alpha}}) -
(\partial_\alpha \overline{\gamma_{t;\alpha}})\,
(\partial_t \gamma_{t;\alpha}) = 2i\, 
|f_t'(u)|^2\, \rho_t(u)
\label{bdry}
\end{eqnarray}

Hence, the evolution of the domain may be encoded either in the
evolution law for its uniformizing conformal map as in eq.(\ref{loew}) 
or in the boundary normal velocity as in eq.(\ref{bdry}).
The two equations are equivalent.

\subsection{Examples}

\subsubsection{Stochastic Loewner evolution (SLE)} 
>From this point of view, SLE is singular as it corresponds to a Dirac
measure $\delta_{U_t}(u)$ centered at the position of a Brownian
motion on the unit circle. The locality of this measure reflects the
fact that SLE describes growing curves. Continuity of the Brownian
motion reflects the absence of branching in the SLE growths. To make
contact with the previous sections, the reader should check that
eq.(\ref{eq:sler}) becomes of the form (\ref{loewbis}) after the
cylinder has been mapped to the outside of the unit disk. 

\subsubsection{Laplacian growth (LG)}
This is a process in which the growth of the domain is governed by the
solution of Laplace equation, i.e. by an harmonic function, in the
exterior of the domain with appropriate boundary conditions. It
originates from the hydrodynamical Hele-Shaw problem to be described
below.

To be precise, let $P$ be the real solution of Laplace equation,
$\nabla^2P=0$, in $\mathbb{O}_t$ with the boundary behavior $P=-\log
|z| + \cdots$ at infinity and $P=0$ on the boundary curve
$\Gamma_t=\partial \mathbb{O}_t$. The time evolution of the domain is
then defined by demanding that the normal velocity of points on the
boundary curve be equal to minus the gradient of $P$: $v_n=-(\nabla
P)_n$.

This problem may be written as a Loewner chain since, as is well
known, Laplace equation is solved via complex analysis by writing $P$
as the real part of an analytic function.  One first solves Laplace
equation in the complement of the unit disk with the appropriate
boundary conditions and then transports it back to the physical domain
$\mathbb{O}_t$ using the map $f_t$. This gives:
\begin{eqnarray}
P= -\Re{\rm e}\ \Phi_t\quad {\rm with}\quad 
\Phi_t(z) = \log g_t(z) 
\nonumber
\end{eqnarray}

The evolution equation for the map $f_t$ is derived using 
that the boundary normal velocity is $v_n=-(\nabla P)_n$. 
The above expression for $P$ gives:
$$ 
v_n = -(\nabla P)_n = |f_t'(u)|^{-1}
$$
at point $\gamma_t=f_t(u)$ on the boundary curve.  As explained in the
previous section, this is enough to determine $\partial_t f_t(w)$ for
any $|w|>1$ since this data specifies the real part on the unit circle
of the analytic function $\partial_tf_t(w)/wf_t'(w)$ on the complement
of the unit disk. The result is:
\begin{eqnarray}
\partial_t f_t(w) = wf_t'(w)\,\oint_{|u|=1}
\frac{du}{2i\pi u\, |f_t'(u)|^2} 
\Big({\frac{w+u}{w-u}}\Big)
\label{lg}
\end{eqnarray}
It is a Loewner chain with $\rho_t(u)=|f_t'(u)|^{-2}$.  

As we shall see below, Laplacian growth is an integrable system, which
may be solved exactly, but it is ill-posed as the domain develops
singularities (cusps $y^2\simeq x^3$) in finite time.  It thus needs
to be regularized. There exist different ways of regularizing it.

One may also formulate Laplacian growth using a language borrowed from
electrostatics by imagining that the inner domain is a perfect
conductor. Then $V=\Re{\rm e}\,\Phi_t$ is the electric potential which
vanishes on the conductor but with a charge at infinity. The electric
field $\vec{E}=\vec{\nabla}V$ is $\bar E\equiv E_x-i
E_y=\partial_z\,\Phi_t$. Its normal component
$E_n=|f_t'(u)|^{-1}$ is proportional to the surface charge density.

\subsubsection{The Hele-Shaw problem (HS)}
This provides a hydrodynamic regularization of Laplacian growth.
The differences with Laplacian growth are in the boundary conditions
which now involve a term proportional to the surface tension.
It  may be formulated as follows \cite{heleshaw}. 

One imagines that the domain $\mathbb{K}_t$ is filled with a non
viscous fluid, say air, and the domain $\mathbb{O}_t$ with a viscous
one, say oil. Air is supposed to be injected at the origin and there
is an oil drain at infinity. The pressure in the air domain
$\mathbb{K}_t$ is constant and set to zero by convention.  In
$\mathbb{O}_t$ the pressure satisfies the Laplace equation
$\nabla^2P=0$ with boundary behavior $P=-\phi_\infty \log |z| +
\cdots$ at infinity reflecting the presence of the oil drain.  The
boundary conditions on the boundary curve are now $P=-\sigma\kappa_t$
with $\sigma$ the surface tension and $\kappa_t$ the curvature of the
boundary curve~\footnote{The curvature is defined by $\kappa\equiv
  -{\vec{n}.\partial_s\vec{\tau}}/{\vec{\tau}^2} = \Im{\rm
    m}[{\bar\tau\partial_s\tau}/{|\tau|^3}]$ with $\vec{\tau}$ the
  tangent and $\vec{n}$ the normal vectors. An alternative formula is:
  $\kappa=|f_t'(u)|^{-1}\Re{\rm e} [1+\frac{uf_t''(u)}{f_t'(u)}]$. For
  a disk of radius $R$, the curvature is $+1/R$.}.  The fluid velocity
in the oil domain $\mathbb{O}_t$ is $\vec{v}=-\vec{\nabla}P$. Laplace
equation for $P$ is just a consequence of incompressibility. The
evolution of the shape of the domain is specified by imposing that
this relation holds on the boundary so that the boundary normal
velocity is $v_n=-(\nabla P)_n$ as in Laplacian growth.

Compared to Laplacian growth, the only modification is the boundary
condition on the boundary curve. This term prevents the formation of
cusps with infinite curvature singularities. The parameter
$\phi_\infty$ sets the scale of the velocity at infinity.
In the following we set $\phi_\infty=1$. By dimensional analysis
this implies that $[{\tt time}]$ scales as $[{\tt length}^2]$
and the surface tension $\sigma$ has dimension of a $[{\tt length}]$.
It plays the role of an ultraviolet cut-off.

A standard procedure \cite{heleshaw} to solve the equations for the
Hele-Shaw problem is by first determining the pressure using complex
analysis and then computing the boundary normal velocity.  By Laplace
equation, the pressure is the real part of an analytic function, $P=
-\Re{\rm e} \Phi_t$. The complex velocity $v=v_x+iv_y$ is $\bar v=
\partial_z\Phi_t$. At infinity $\Phi_t(z)\simeq \log z+\cdots$ and
$\bar v\simeq 1/z+\cdots$.  The boundary conditions on $P$ demand that
$$
(\Phi_t\circ f_t)(w) = \log w + \sigma \vartheta_t(w)
$$ 
where $\vartheta_t(w)$ is analytic in
$\mathbb{D}$ with boundary value $\Re{\rm e}[\vartheta_t(u)]=
\kappa_t(f_t(u))$ with $\kappa_t$ the curvature.  Explicitly
$$
\vartheta_t(w)=\oint\frac{du}{2i\pi u}\,
\Big(\frac{w+u}{w-u}\Big)\,\kappa_t(f_t(u))
$$  
The evolution of $f_t$ is then found by evaluating the boundary normal 
velocity $v_n=\Re{\rm e}(\nabla\Phi)_n$ at point $\gamma_t=f_t(u)$: 
$$ 
v_n = \Re{\rm e}[\, n\partial_z\Phi_t\,]
= |f_t'(u)|^{-1}\, \Re{\rm e}[1+\sigma u \partial_u\theta_t(u)]
$$
As above, this determines uniquely
$\partial_tf_t(u)$ and it leads to a Loewner chain (\ref{loew})
with density:
\begin{eqnarray}
\rho_t(u) = |f_t'(u)|^{-2}\, \Big( 1 + \sigma \epsilon_t(u)\Big)
\quad,\quad \epsilon_t(u) = \Re{\rm e}[ u\partial_u \vartheta_t(u)]
\label{meshLG}
\end{eqnarray}

The difference with Laplacian growth is in the extra term
proportional to $\sigma$. It is highly non-linear and non-local. 
This problem is believed to be well defined at all times for $\sigma$
positive.  

\subsubsection{Other regularized Laplacian growth (rLG)}

These regularizations amount to introduce an UV cutoff $\delta$ in the
physical space by evaluating $|f_t'|$ at a finite distance away from
$\partial\mathbb{O}_t$. A possible choice \cite{MakCarl} is
$\rho_t(u)^{1/2}=\delta^{-1}{\rm inf}\{\varepsilon:\, {\rm
  dist}[f_t(u+\varepsilon u);\partial\mathbb{O}_t]=\delta\}$.  An
estimation gives $\rho_t(u)\asymp |f_t'(u+\hat\varepsilon_u u)|^{-2}$
where $\hat\varepsilon_u$ goes to $0$ with $\delta$, so that it naively
approaches $|f_t'(u)|^{-2}$ as $\delta\to 0$.  

Another possible, but less physical, regularization consists in
introducing an UV cutoff $\nu$ in the mathematical space so that
$\rho_t(u)=|f_t'(u+\nu u)|^{-2}$.

\subsubsection{Dielectric breakdown and generalizations}
A larger class of problems generalizing Laplacian growth have been
introduced. Their Loewner measures are as in Laplacian growth but with a
different exponent:
$$\rho_t(u)=|f_t'(u)|^{-\alpha}\quad,\quad 0\leq\alpha\leq2.$$
Using an
electrostatic interpretation of the harmonic potential, one usually
refers to the case $\alpha=1$ as a model of dielectric breakdown
because the measure is then proportional to the local electric field
$E_n=|f_t'(u)|^{-1}$. This is a phenomenological description. Just as
the naive Laplacian growth these models are certainly ill-posed.  They
also require ultraviolet regularizations, one of which is described
below using iterated conformal maps.

\subsection{Cusp singularities in LG}

The naive LG problem, without regularization, corresponds to the
Loewner density $\rho_t(u)=|f_t'(u)|^{-2}$.  The occurrence of
singularities may be grasped by looking for the evolution of domains
with a $Z_n$ symmetry uniformized by the maps
$$
f_t(w)=R_tw(1+\frac{\beta_t}{n-1}w^{-n})
$$
for some $n>2$ and with $|\beta_t|\leq 1$. This form of 
conformal maps is preserved by the dynamics. The conformal radius
$R_t$ and the coefficients $\beta_t$ evolve with time according to
$\partial_t R^2_t= 2/(1-\beta_t^2)$ and $\beta_t = (R_t/R_c)^{n-2}$ with
$R_c$ some integration constant.  The singularity appears when
$\beta_t$ touches the unit circle which arises at a finite time $t_c$.
At that time the conformal radius is $R_c$.

\begin{figure}[htbp]
\begin{center}
\includegraphics[width=0.9\textwidth]{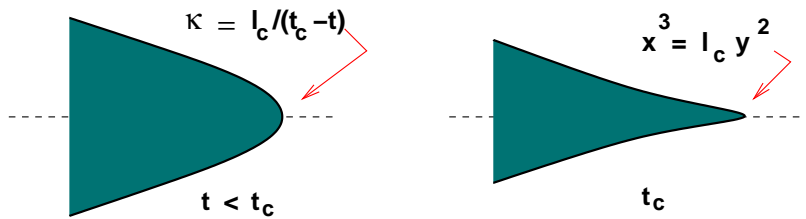}
\caption{Cups formation in Laplacian growth.}
     \label{fig:cusp}
 \end{center}
\end{figure}

At $t_c$ the boundary curve $\Gamma_{t_c}$ has
cusp singularities of the generic local form 
$$
\ell_c\,(\delta y)^2\simeq (\delta x)^3
$$
with $\ell_c$ a characteristic local length scale.
In the present simple case $\ell_c\simeq R_c$. 
At time $t\nearrow t_c$, the dynamics is regular in the
dimensionless parameter $\ell_c^{-1}\sqrt{t_c-t}$. 
The maximum curvature of the boundary curve scales
as $\kappa_{\rm max} \simeq {\ell_c}/{(t_c-t)}$ near $t_c$ and it is
localized at a distance $\sqrt{t_c-t}$ away from the would be cusp
tip. See Figure (\ref{fig:cusp}).

This behavior is quite generic. As we shall see, conformal maps
$f_t(w)$ such that their derivatives are polynomials in $w^{-1}$ are
stable by the Laplacian growth dynamics. By construction, their zeroes are
localized inside the unit disk. A singularity in the boundary
curve occurs if one of these zeroes converges to the unit circle.
The singularity is then a cusp   $\ell_c\,y^2\simeq x^3$ as can be
seen by expanding locally the conformal map around its singular point.

Once regularized with an explicit ultraviolet cut-off, the processes
are believed to be well defined for all time. The effect of the
regularization procedure on the domain properties is presently
unclear. The domain structures may a priori depend on how the problem
has been regularized. In the hydrodynamic regularization --the
Hele-Shaw problem-- the cusp production is expected to be replaced
by unlimited ramifications leading to dendritic growth.

In the regularized model, the curvature of $\Gamma_t$ is expected to
remain finite at all time. Using scaling theory, a crude estimate of
its maximum around the would be singularities may be obtained by
interchanging the short distance scale $\sqrt{(t_c-t)}$ near the
singularity in the unregularized theory with the UV cutoff of the
regularized theory.  In the hydrodynamic regularization this
gives $\kappa_{\rm max}\simeq \ell_c/\sigma^2$ as $\sigma\to 0$.

\subsection{Discrete Loewner chains}

\subsubsection{DLA}

DLA stands for diffusion limited aggregation \cite{dla}. It refers to
processes in which the domains grow by aggregating diffusing
particles. Namely, one imagines building up a domain by clustering
particles one by one. These particles are released from the point at
infinity, or uniformly from a large circle around infinity, and
diffuse as random walkers. They will eventually hit the domain and the
first time this happens they stick to it. By convention, time is
incremented by unity each time a particle is added to the domain. Thus
at each time step the area of the domain is increased by the physical
size of the particle. The position at which the particle is added
depends on the probability for a random walker to visit the boundary
for the first time at this position.

\begin{figure}[htbp]
\begin{center}
\includegraphics[width=0.6\textwidth]{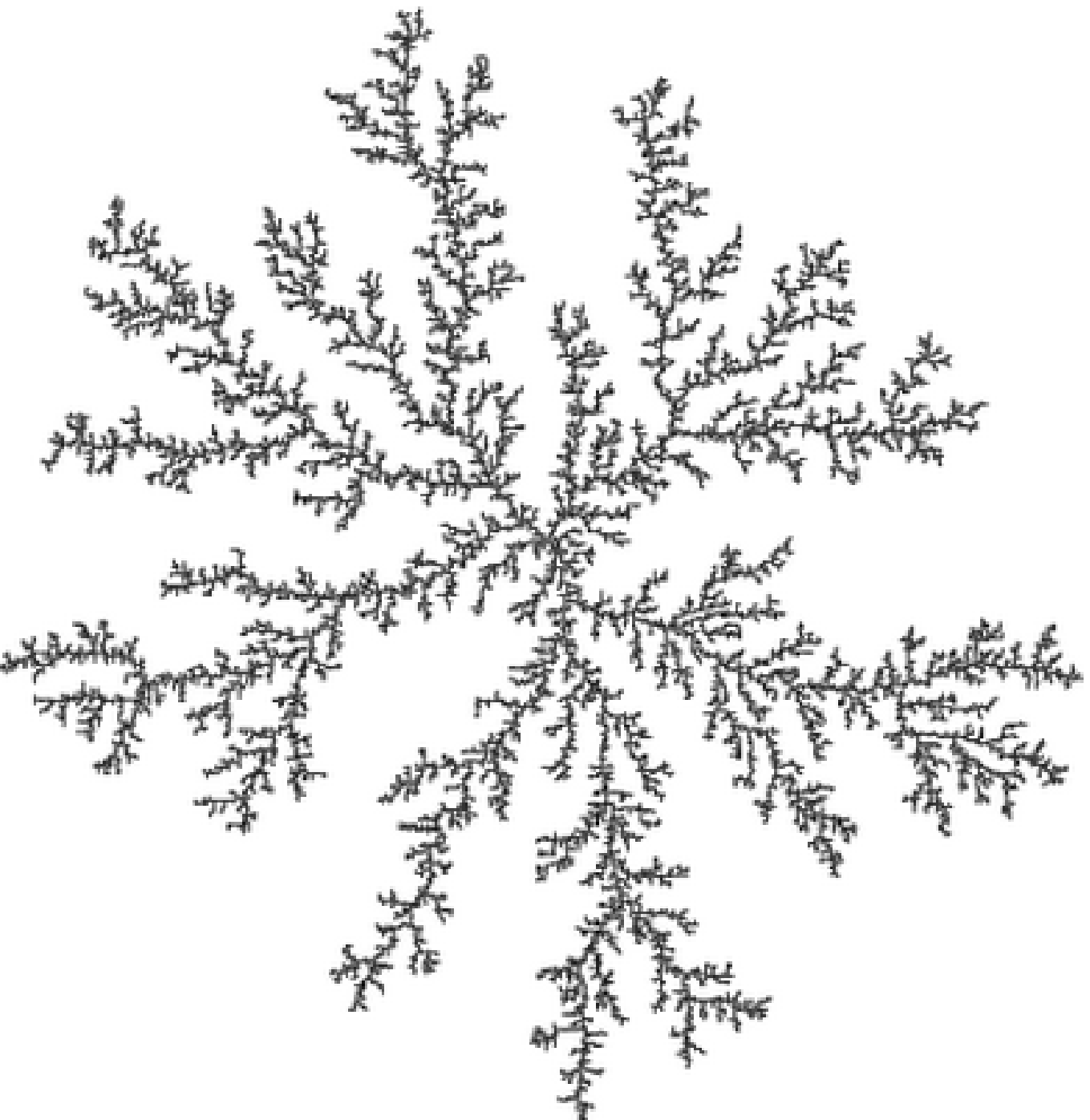}
\caption{A DLA sample.}
     \label{fig:dla}
 \end{center}
\end{figure}

In a discrete approach one may imagine that the particles are tiny
squares whose centers move on a square lattice whose edge lengths
equal that of the particles, so that particles fill the lattice when
they are glued together. The center of a particle moves as a random
walker on the square lattice. The probability $Q(x)$ that a particle
visits a given site $x$ of the lattice satisfies the lattice version
of the Laplace equation $\nabla^2Q=0$.  It vanishes on the boundary of
the domain, i.e. $Q=0$ on the boundary, because the probability for a
particle to visit a point of the lattice already occupied, i.e. a
point of the growing cluster, is zero.  The local speed at which the
domain is growing is proportional to the probability for a site next
to the interface but on the outer domain to be visited. This
probability is proportional to the discrete normal gradient of $Q$,
since the visiting probability vanishes on the interface. So the local
speed is $v_n=(\nabla Q)_n$. It is not so easy to make an unbiased
simulation of DLA on the lattice. One of the reasons is that on the
lattice there is no such simple boundary as a circle, for which the
hitting distribution from infinity is uniform. The hitting
distribution on the boundary of a square is not such a simple
function. Another reason is that despite the fact that the symmetric
random walk is recurrent is 2d, each walk takes many steps to
glue to the growing domain. The typical time to generate a single
sample of reasonable size with an acceptable bias is comparable to the
time it takes to make enough statistics on loop-erased random walks or
percolation to get the scaling exponent with two significant digits.
Still this is a modest time, but it is enough to reveal the intricacy
of the patterns that are formed. Fig.\ref{fig:dla} is such a
sample. The similarity with the sample in fig.\ref{fig:dlaconf},
obtained by iteration of conformal maps, is striking. But a
quantitative comparison of the two models is well out of analytic
control and belongs to the realm of extensive simulations.

DLA provides a discrete analogue of Laplacian growth. The particle
size plays the role of an ultraviolet cutoff. Since DLA only assumes
that the growth is governed by the diffusion of elementary patterns,
DLA has been applied to a large variety of aggregation or deposition
phenomena, see eg.\cite{revue}.

During this process the clustering domain gets ramified and develops
branches and fjords of various scales. The probability for a particle
to stick on the cluster is much higher on the tip of the branches than
deep inside the fjords. This property, relevant at all scales, is
responsible for the fractal structure of the DLA clusters.
 
Since its original presentation \cite{dla}, DLA has been studied
numerically quite extensively. There is now a consensus that the
fractal dimension of 2d DLA clusters is $D_{\rm dla}\simeq 1.71$.
There is actually a debate on whether this dimension is geometry
dependent but a recent study \cite{channel} seems to indicate that DLA
clusters in a radial geometry and a channel geometry have identical
fractal dimension. To add a new particle to the growing domain, a
random walk has to wander around and the position at which it finally
sticks is influenced by the whole domain. To rephrase this, for each
new particle one has to solve the outer Laplace equation, a non-local
problem, to know the sticking probability distribution. This is a
typical example when scale invariance is not expected to imply
conformal invariance.

\subsubsection{Iterated conformal maps}

As proposed in \cite{lgdiscret}, an alternative way to mimic the
gluing of elementary particles consists in composing elementary
conformal maps, each of which corresponds to adding an elementary
particle to the domain.

One starts with an elementary map corresponding to the gluing of a
tiny bump, of linear size $\lambda$, to the unit disk. A large variety
of choices is possible, whose influence on the final structure of the
domain is unclear. An example is given by the following formul\ae\
($g_\lambda$ is the inverse map of $f_\lambda$):
\begin{eqnarray}
g_\lambda (z) &=& z\  \frac{ z\cos\lambda-1}{z-\cos\lambda}\nonumber\\
f_\lambda (w)= &=&
(2\cos\lambda)^{-1}\left[{w+1+\sqrt{w^2-2w\cos2\lambda
      +1}}\right]\nonumber 
\end{eqnarray}
where $f_\lambda$ correspond to the deformation of the unit disk
obtained by gluing a semi-disk centered at point $1$ and whose two
intersecting points with the unit circle define a cone of angle
$2\lambda$. For $\lambda\ll 1$, the area of the added bump is of order
$\lambda^2$.  But other choices are possible and have been used.

Gluing a bump around point $e^{i\theta}$ on the unit circle is
obtained by rotating  these maps. The uniformizing maps are then
$$ 
f_{\lambda;\theta}(w)= e^{i\theta}\,f_{\lambda}(we^{-i\theta})
$$

\begin{figure}[htbp]
\begin{center}
\includegraphics[width=0.9\textwidth]{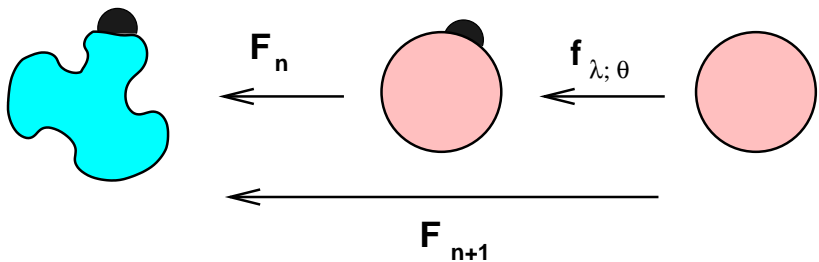}
\caption{Iteration of conformal maps.}
     \label{fig:iter}
 \end{center}
\end{figure}

The growth of the domain is obtained by successively iterating the maps
$f_{\lambda_n;\theta_n}$ with various values for the size
$\lambda_n$ and the position $\theta_n$ of the bumps. 
See Figure (\ref{fig:iter}).
Namely, if after $n$ iterations the complement of the unit disk is
uniformized into the complement of  the domain by the map
$F_{(n)}(w)$, then at the next $(n+1)^{\rm th}$ iteration the
uniformizing map is given by:
\begin{eqnarray}
F_{(n+1)}(w)= F_{(n)}(\, f_{\lambda_{n+1};\theta_{n+1}}(w)\, )
\label{iter}
\end{eqnarray}
For the inverse maps, this becomes
$G_{(n+1)}=g_{\lambda_{n+1};\theta_{n+1}}\circ G_{(n)}$.

%\vskip 1.0 truecm

To fully define the model one has to specify the choice of the
parameter $\lambda_n$ and $\theta_n$ at each iteration. 
Since $\lambda_n$ codes for the linear size of the added bump and since
locally conformal maps act as dilatations, the usual choice is to
rescale $\lambda_{n+1}$ by a power of $|F_{(n)}'(e^{i\theta_n})|$ as:
$$ 
\lambda_{n+1} = \lambda_0\ |F_{(n)}'(e^{i\theta_n})|^{-\alpha/2},
\quad 0\leq \alpha \leq 2
$$
The case $\alpha=2$ corresponds to DLA as the physical area of the
added bump are approximatively constant and equal to $\lambda_0$ at
each iterations.  In the other case, the area of the added
bump scales as $|F_{(n)}'(e^{i\theta_n})|^{2-\alpha}$.

The positions of the added bump are usually taken uniformly
distributed on the mathematical unit circle with a measure
$d\theta/2\pi$.

It is clear that this discrete model with $\alpha=2$ provides a
regularization of Laplacian growth with $\lambda_0$ playing the role
of an ultraviolet cutoff. This may also be seen by looking at the {\it
  naive} limit of a small cutoff. Indeed, a naive expansion as
$\lambda_n\ll 1$ gives that $F_{(n+1)}=F_{(n)}+\delta F_{(n)}$ with
$$
\delta F_{(n)}(w)\simeq \lambda_n\, wF_{(n)}' (w)\,
\left(\frac{w+e^{i\theta_n}}{w-e^{i\theta_n}}\right)
$$
where we used the expression of $f_{\lambda}$ for $\lambda\ll 1$.
Using the recursive expression for $\lambda_n$ and averaging over
$\theta$ with a uniform distribution yields:
$$
\langle \delta F(w) \rangle = \lambda_0\, w F'(w)\,
\oint \frac{d\theta}{2\pi} |F'(e^{i\theta})|^{-\alpha}\,
\frac{w+e^{i\theta}}{w-e^{i\theta}}
$$
For $\alpha=2$ this reproduces the Loewner chain for Laplacian
growth.  But this computation is too naive as the small cutoff limit
is not smooth, a fact which is at the origin of the non trivial
fractal dimensions of the growing domains.

There are only very few mathematical results on these discrete models.
The most recent one \cite{rhode} deals with the simplest (yet
interesting but not very physical) model with $\alpha=0$. It proves the
convergence of the iteration to well-defined random maps uniformizing
domains of Hausdorff dimension $1$. However, these models have been
studied numerically extensively. There exists a huge literature on
this subject but see ref.\cite{discret} for instance. These studies
confirm that the fractal dimension of DLA clusters with $\alpha=2$ is
$D_{\rm dla}\simeq 1.71$ but they also provide further informations on
the harmonic measure multi-fractal spectrum.  Results on the $\alpha$
dependence of the fractal dimension may be found eg. in
ref.\cite{hastbis}.

Various generalizations have been introduced. For instance, in
ref.\cite{gener} a model of iterated conformal maps has been defined
in which particles are not added one by one but by layers.  These
models have one control parameter coding for the degree of coverage of
the layer at each iterative step. By varying this parameter the
model interpolates between discrete DLA and a discrete version of the
Hele-Shaw problem. The fractal dimension of the resulting clusters varies
with this parameter and seems to indicate that the fractal dimension
of this discrete analogue of the Hele-Shaw problem is $2$, a point
that we shall discuss further in the last section.

\subsection{Integrability of Laplacian growth}

Laplacian growth is an integrable system, at least up to the cusp
formation. Let us recall that it corresponds to a Loewner chain with a 
density $\rho_t(u)=|f_t(u)|^{-2}$, or equivalently to the quadratic
equation
\begin{eqnarray}
(\partial_\alpha \gamma_{t;\alpha})\,
(\partial_t \overline{\gamma_{t;\alpha}}) -
(\partial_\alpha \overline{\gamma_{t;\alpha}})\,
(\partial_t \gamma_{t;\alpha}) = 2i\,
\label{lgbdry}
\end{eqnarray}
for the dynamics of the boundary points $\gamma_{t;\alpha}=f_t(u)$,
$u=e^{i\alpha}$. What makes the model integrable is the fact that the
r.h.s of eq.(\ref{lgbdry}) is constant. Eq.(\ref{lgbdry}) is then
similar to a quadratic Hirota equation.  Hints on the integrable
structure were found in \cite{shraiben} and much further developed in
\cite{richards,integr}.

\subsubsection{Conserved quantities}

We now define an infinite set of quantities which are conserved in the
naive unregularized LG problem. They reflect its integrability. We
follow ref.\cite{richards,integr}.  These quantities may be defined via a
Riemann-Hilbert problem on $\Gamma_t$ specified by,
 \begin{eqnarray}
S_+(\gamma)-S_-(\gamma) = \bar \gamma \quad,\quad \gamma\in \Gamma_t
\label{lgrh}
\end{eqnarray} 
for functions $S_-$ and $S_+$ respectively analytic in the outer
domain $\mathbb{O}_t$ and in the inner domain $\mathbb{K}_t$.
We fix normalization by demanding $S_-(\infty)=0$.
We assume $\Gamma_t$ regular enough for this
Riemann-Hilbert problem to be well defined. 
As usual, $S_\pm$ may be presented as contour integrals:
$$
S_\pm(z)=-\oint_{\Gamma_t}\frac{d\gamma}{2i\pi} \frac{\bar
  \gamma}{z-\gamma}.
$$ 

The conserved quantities are going to be expressed in terms of
$S_\pm$. We thus need their time evolution.  Differentiation of
eq.(\ref{lgrh}) with respect to time and use of the evolution
equation (\ref{lgbdry}) gives:
$$
 \partial_t S_+(\gamma)-\partial_t S_-(\gamma) = 
2{g_t'(\gamma)}/{g_t(\gamma)} 
$$
Notice now that ${g_t'(\gamma)}/{g_t(\gamma)}$ is the boundary
value of $(\log g_t)'$ which by construction is analytic in
$\mathbb{O}_t$. We may thus rewrite this equation as a trivial
Riemann-Hilbert problem, $ \partial_t S_+(\gamma)-(\partial_t S_- +
2(\log g_t)')(\gamma) =0$, so that both terms vanish:
\begin{eqnarray}
\partial_tS_+(z)=0
\quad {\rm and}\quad
(\partial_t S_- + 2(\log g_t)')(z)=0
\label{scw+-}
\end{eqnarray}

Since $S_+$ is analytic around the origin, we may expand it in
power of $z$.  Equation $\partial_t S_+(z)=0$ then tells us that
$S_+(z)$ is a generating function of conserved quantities:
$S_+(z)=\sum_{k\geq 0}z^{k} I_k$ with 
\begin{eqnarray}
I_k=\oint_{\Gamma_t} \frac{d\gamma}{2i\pi} \bar\gamma \gamma^{-k-1}
\quad,\quad \partial_t I_k=0. 
\label{lgik}
\end{eqnarray}
This provides an infinite set of conserved quantities.

Since $S_-$ is analytic around infinity, it may be expanded in power
of $1/z$: $S_-(z)=-\mathcal{A}_t/\pi z+\cdots$ with
$\mathcal{A}_t=-\frac{i}{2}\oint_{\Gamma_t}d\gamma \bar \gamma$ the
area of $\mathbb{K}_t$.  The second equation $(\partial_t S_- + 2(\log
g_t)')(z)=0$ with $g_t(z)=R_t^{-1}z+O(1)$ then implies
$\partial_t\mathcal{A}_t=2\pi.$
The area of the domain grows linearly with time, up to the time
at which the first cusp singularity appears.

\subsubsection{Simple solutions}

A particularly simple class of conformal maps, solutions of the
Laplacian growth equation, are those such that their derivatives are
polynomials in $w^{-1}$. They may be expanded as:
\begin{eqnarray}
f_t(w)=\sum_{n=0}^N f_nw^{1-n},\quad f_0=R_t>0
\label{poly}
\end{eqnarray}
with $N$ finite but arbitrary. The dynamical variables are the $N+1$
coefficient $f_0,\cdots, f_N$. 
They are all complex except $f_0$ which is real.
It will be convenient to define the function $\bar f_t$ by
$\bar f_t(w)=\overline{f_t(\overline{w})}$.

The fact that this class is stable under the dynamics follows from the
Loewner equation (\ref{lg}). The trick consists in using the fact that
the integration contour is on the unit circle so that 
$|f_t'(u)|^2=f_t'(u)\bar f_t'(1/u)$. The contour integral then
involves a meromorphic function of $u$ so that it can be evaluated by
deforming the contour to pick the residues. This is enough to prove
that $\partial_tf_t(w)$ possesses the same structure as $f_t(w)$
itself so that the class of functions (\ref{poly}) is stable under the
dynamics.

Alternatively one may expand the quadratic equation
(\ref{lgbdry}) to get a hierarchy of equations:
$$
\sum_{n\geq 0} (1-n)[f_n\dot{\bar f}_{j+n} + \bar f_n\dot f_{-j+n}]
=2\delta_{j;0}
$$
For $j=0$, this equation tells us again that the area of the domain grows
linearly with time. Besides this relation there are only $N$
independent complex equations for $j=1,\cdots, N$ which actually 
code for the conserved quantities.

To really have an integrable system we need to have as many
independent integrals of motion as dynamical variables. Thus we need
to have $N$ conserved quantities. These are given by the $I_k$'s
defined above which may rewritten as
$$
I_k=\oint_{|u|=1} \frac{du}{2i\pi} 
\frac{f_t'(u)\bar f_t(1/u)}{f_t(u)^{k+1}}
$$
Only the first $N$ quantities, $I_0,\cdots, I_{N-1}=R^{1-N}\bar f_N $
are non-vanishing. They are independent.
They can be used to express algebraically all $f_n$'s, $n\geq 0$, 
in terms of the real parameter $f_0=R_t$. The area law,
$$ 
{\cal A}_t=\pi[R_t^2+\sum_{n\geq 1}(1-n)|f_n|^2]=2\pi t,
$$
with the $f_n$'s expressed in terms of $R_t$, then reintroduces
the time variable by giving its relation with the conformal radius.

\subsubsection{Algebraic curves}

As was pointed out in \cite{integr}, solutions of Laplacian growth and
their cusp formations have an elegant geometrical interpretation
involving Riemann surfaces.

Recall that given a sufficiently smooth real curve $\Gamma_t$ drawn on
the complex plane one may define a function $S(z)$, called the
Schwarz function, analytic in a ribbon enveloping the curve such that
$$ 
S(\gamma) = \overline{\gamma},\quad \gamma\in \Gamma_t
$$
By construction, the Schwarz function may be expressed in terms of
uniformizing maps of the domain bounded by the curves as $S(z)=\bar
f_t(1/g_t(z))$.

The Riemann-Hilbert problem (\ref{lgrh}) defining the conserved
charges then possesses a very simple interpretation: $S_\pm$ are the
polar part of the Schwarz function respectively analytic inside or
outside $\Gamma_t$, i.e. $S(z) = S_+(z) - S_-(z).$ Thus the polar part
$S_+$, analytic in the inner domain, is conserved.  The polar part
$S_-$, analytic in the outer domain, evolves according to
eqs.(\ref{scw+-}). Since $\log g_t(z)$ is analytic in the outer
domain, these equations are equivalent to the single equation:
\begin{eqnarray}
\partial_t S(z) = -2 (\log g_t(z))'
\label{eqSZ}
\end{eqnarray}

Now the physical curve $\Gamma_t$ may be viewed as a real slice of a
complex curve, alias a Riemann surface. The latter is constructed
using the Schwarz function as follows. Recall that $s=S(z)$ is
implicitly defined by the relations $z=f_t(w)$, $s=\bar f_t(1/w)$. In
the case of polynomial uniformizing maps we get the pair of equations
\begin{eqnarray}
z&=& f_0w+f_1+f_2w^{-1}+\cdots +f_Nw^{1-N}\nonumber\\
s&=& \bar f_0w^{-1}+\bar f_1 +\bar f_2 w+\cdots +\bar f_Nw^{N-1}
\nonumber 
\end{eqnarray}
Eliminating $w$ yields an algebraic equation for $z$ and $s$ only:
\begin{eqnarray}
{\bf R}:\quad R(z,s)=0
\label{lgrie}
\end{eqnarray}
with $R$ a polynomial of degree $N$ in both variables, $R(z,s)=\bar
f_N z^N+ f_N s^N+\cdots$.  Eq.(\ref{lgrie}) defines an algebraic curve
${\bf R}$. It is of genus zero since by construction it is uniformized
by points $w$ of the complex sphere. It has many singularities which
have to be resolved to recover a smooth complex manifold.

The Riemann surface ${\bf R}$ may be viewed as a $N$-sheeted covering
of the complex $z$ plane: each sheet corresponds to a determination of
$s$ above point $z$. At infinity, the physical sheet corresponds to
$z\simeq f_0w$ with $w\to\infty$ so that $s\simeq (z/f_0)^{N-1}\,\bar
f_N$, the other $N-1$ sheets are ramified and correspond to $z\simeq
f_N/w^{N-1}$ and $s\simeq \bar f_0/w$ with $w\to 0$ so that $z\simeq
(s/\bar f_0)^{N-1}\, f_N$. Hence infinity is a branch point of order
$N-1$.

By the Riemann-Hurwitz formula the genus $g$ is $2g-2=-2N+\nu$ with
$\nu$ the branching index of the covering. Since the point at infinity
counts for $\nu_\infty=N-2$, there should be $N$ other branch
points generically of order two.  By definition they are determined by
solving the equations $R(z,s)=0$ and $\partial_s R(z,s)=0$. Since the
curve is uniformized by $w\in\mathbb{C}$, these two equations imply
that $z'(w)\partial_zR(z(w),s(w))=0$. Hence either $z'(w)=0$,
$\partial_zR\not=0$, and the point is a branch point, or
$z'(w)\not=0$, $\partial_zR=0=\partial_sR$, and the point is
actually a singular point which needs to be desingularized. So the $N$
branch points at finite distance are the critical points of
the uniformizing map $z=f_t(w)$.

The curve ${\bf R}$ possesses an involution
$(z,s)\to (\bar s,\bar z)$ since $R(\bar s, \bar
z)=\overline{R(z,s)}$ by construction. The set of points fixed by this
involution has two components: (i) a continuous one parametrized by
points $w=u$, $|u|=1$ --this is the real curve $\Gamma_t$ that we
started with-- and (ii) a set of $N$ isolated points which are actually
singular points.

The cusp singularity of the real curve $\Gamma_t$ arises when 
a isolated real point merges with the continuous real slice
$\Gamma_t$. Locally the behavior is as for the curve
$u^2=\varepsilon\, v^2 + v^3$ with $\varepsilon\to 0$.

\begin{figure}[htbp]
\begin{center}
\includegraphics[width=0.8\textwidth]{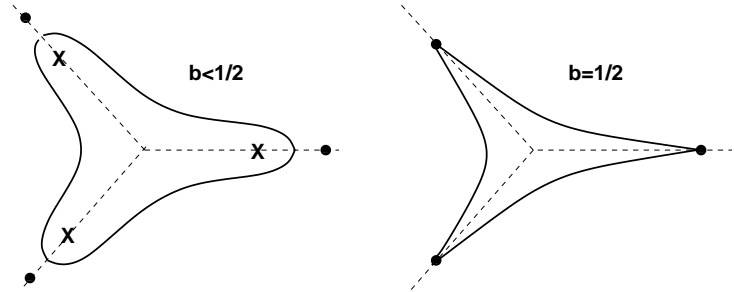}
\caption{Subcritical and critical algebraic curves. Black circles are
  singular points. Crosses are branch points.}
     \label{fig:curve}
 \end{center}
\end{figure}

The simplest example is for $N=3$ with $\mathbb{Z}_3$ symmetry so that
$f_t(w)=w+b/w^2$ and
$$ 
w^2\, z= w^3 +b\quad,\quad w\, s=1+bw^3
$$
We set $f_0=1$ and $f_3=b$.
Without lost of generality we assume $b$ real.
The algebraic curve is then
$$
R(z,s)\equiv bz^3+bs^3 -b^2 s^2z^2 + (b^2-1)(2b^2+1)sz -(b^2-1)^3=0
$$
Infinity is a branch point of order two.  The three other
branch points are at $z=3\omega\,(b/4)^{1/3}$,
$s=\omega^2\,(2b^2+1)(2b)^{-1/3}$ corresponding to
$w=\omega(2b)^{1/3}$ with $\omega$ a third root of unity. They are
critical points of $z(w)$.  There are three singular points at
$z=\omega\,(1-b^2)/b$, $s=\omega^2\, (1-b^2)/b$ corresponding to
$w=\omega(1\pm\sqrt{1-4b^2})/2b$.  The physical regime is for $b<1/2$
in which case the real slice $\Gamma_t=\{z(u),\, |u|=1\}$ is a simple
curve.  The singular points are then in the outer domain and the
branch points in the inner domain. The cusp singularities arise for
$b=1/2$.  For $b>1/2$ there are no isolated singular points, they are
all localized on the real slice so that $\Gamma_t$ possesses double
points. See Figure (\ref{fig:curve}).

%\vskip 1.0 truecm

\subsection{Dendritic anomalies}

We now would like to discuss a few points concerning the regularized
models and their small cut-off limits --which are actually what we are
interested in. We shall point out that the hydrodynamic regularization
used in the Hele-Shaw problem possesses essential differences with
that used in other regularized models, say DLA. This opens the
possibility for the Hele-Shaw problem and DLA not to be in the same
universality class.

This observation is based on the conjectural existence of anomalies in
the Hele-Shaw problem --a word which refers to quantities which, although
naively vanishing  in the small cut-off limit, are actually non zero
but finite in this limit due to compensating effects.

An analogy with Burgers turbulence may be useful. The 1d Burgers
equation is $\partial_tu+u\partial_xu-\nu\partial_x^2u=0$ for a
velocity field $u(t,x)$. The viscosity $\nu$ plays the role of
ultraviolet cut-off. At $\nu=0$ this equation is simply Euler equation
which may be easily solved. At $\nu=0$, any smooth initial data
produces shocks in finite time with a discontinuity of $u$. In the
regularized equation with $\nu\not=0$, there is no shock production
but the field $\hat \epsilon \equiv \nu (\partial_x u)^2$, which
naively vanishes when $\nu\to 0$, has actually a non trivial limit.
It does not vanish because there is a compensation between small
viscosity $\nu$ and large velocity gradient $(\partial_xu)$. The field
$\hat \epsilon$ codes for the amount of energy dissipated inside the
shocks.

The analogy that we would like to put forward is that the cusp
formation in Laplacian growth is analogous to shock formation in
Burgers and that the surface tension $\sigma$ in Laplacian growth plays
the role of the viscosity in Burgers. There are potential compensating
effects between small surface tension and large curvature in the
Hele-Shaw problem.

\subsubsection{Almost conserved quantities}

We now describe the effect of the regularization procedure on the 
conserved quantities of Laplacian growth. Recall that they were defined 
starting from the Riemann-Hilbert problem:
 \begin{eqnarray}
h_+(\gamma)-h_-(\gamma) = \bar \gamma \quad,\quad \gamma\in \Gamma_t
\label{loewrh}
\end{eqnarray} 
for functions $h_-$ and $h_+$ respectively analytic in $\mathbb{O}_t$ and
$\mathbb{K}_t$. We use the notation $h_\pm$ instead of $S_\pm$ in
order to avoid confusion with the unregularized Laplacian growth.  We
again fix normalization by demanding $h_-(\infty)=0$ so that
$h_\pm(z)=-\oint_{\Gamma_t}\frac{d\gamma}{2i\pi} \frac{\bar
  \gamma}{z-\gamma}$.  Time differentiating eq.(\ref{loewrh}) using
the evolution equation (\ref{bdry}) now gives:
\begin{eqnarray}
 \partial_t h_+(\gamma)-\partial_t h_-(\gamma) = 
2\frac{g_t'(\gamma)}{g_t(\gamma)} \frac{v_n(\gamma)}{|g_t'(\gamma)|}
\label{rhbis}
\end{eqnarray}
with $v_n$ the boundary normal velocity --in LG, we had
$v_n=|g_t'(\gamma)|$.

In the case of the Hele-Shaw (HS) problem, the hydrodynamic
regularization of Laplacian growth, the normal velocity possesses a
particular form (\ref{meshLG}) involving the derivative of
$\vartheta$.  Eq.(\ref{rhbis}) may then be rewritten in terms of the
potential, $\Phi_t(z)=\log g_t(z) + \sigma\vartheta_t(g_t(z))$, and of
the curvature as:
$$
\partial_t h_+(\gamma)-\partial_t h_-(\gamma) =
2 \Phi_t'(\gamma) -2\sigma\kappa_t'(\gamma)
$$
Since $\Phi_t$ is analytic in $\mathbb{O}_t$, the solution of this
equation is given by that of the Riemann-Hilbert problem for
$\kappa_t$. So, let $\kappa_{-}$ and $\kappa_{+}$ be respectively
analytic in $\mathbb{O}_t$ and $\mathbb{K}_t$ with boundary values
$\kappa_{+}-\kappa_{-}=\kappa_t$ on $\Gamma_t$. Then:
\begin{eqnarray}
\partial_t h_+(z)&=& - 2\sigma \partial_z\, \kappa_{+}(z)
\label{quasicon}\\
\partial_t h_-(z) &=& -2 \partial_z\Phi_t(z) 
- 2\sigma \partial_z\, \kappa_{-}(z) \nonumber
\end{eqnarray}
These equations codes for the behavior of $h_\pm$ in the limit of
vanishing surface tension.

A noticeable point is that the above r.h.s. are total derivatives.  In
particular, contribution to the $1/z$ term in $\partial_th_-$ only
comes from the logarithmic factor in the potential.  This implies that
at $\sigma\not= 0$ the area of the domains also grows linearly with
time, $\mathcal{A}_t\simeq 2\pi t$ at large time~\footnote{This result
  concerning the area is simply a consequence of the fluid
  incompressibility.  But the general method allows to deal with the
  complete hierarchy of quasi conserved charges.}, and it is
independent of the UV cutoff $\sigma$.  This is contrast with other
regularizations for which the area growth may be cutoff dependent.

\subsubsection{A conjecture}

The supplementary terms proportional to $\sigma\kappa'_\pm$ in the
quasi conservation laws (\ref{quasicon}) naively vanish when the UV
cutoff is removed.  {\it Dendritic anomalies} refer to the possibilities for
these terms not to vanish as the cutoff goes to zero, say $\sigma\to
0$ in the Hele-Shaw (HS) problem.  Such possibility arises due to a
balance between small surface tension and large curvature.

Recall the previous HS estimation of the maximum curvature $\kappa_{\rm
  max} \simeq \ell_c/\sigma^2$ with $\ell_c$ the local scale
associated to the cusp.  These maxima are localized around
points $\gamma_c$ which would turn into cusp singularities in
absence of surface tension.  So, we may expect that in the HS problem
the product $\sigma\kappa(\gamma)$ is a function picked around its
maximum values and is of the form
$\frac{\ell_c}{\sigma}\varphi(\frac{\gamma-\gamma_c}{\sigma})$ when the
point $\gamma$ moves away from $\gamma_c$ with $\varphi$ finite at the
origin and decreasing rapidly away from it. For isolated singular
points $\gamma_c$, this would converge in the limit $\sigma\to 0$ to a
sum of Dirac point measures.

This leads us to conjecture that the anomalous term $\sigma
\kappa(\gamma)$ goes to a non trivial but finite distribution as
$\sigma\to 0$.  This distribution then contributes to
eq.(\ref{quasicon}) to break the `classical' conservation laws valid
in the unregularized theory, hence the name {\it dendritic anomaly}.

The Loewner equation (\ref{loew}) --or the evolution equation
(\ref{bdry})-- are then expected to have a finite limit as $\sigma\to
0$ with a non trivial right hand side. The solution would be
well-defined for all times and would describe a non trivial familly
$\hat \mathbb{K}_t$ such that, for fixed $t$, $\hat \mathbb{K}_t$ is
the limit of the domains $\mathbb{K}_t$ as $\sigma\to 0$. This is
consistent with the observation that the area of $ \mathbb{K}_t$ is
independant of $\sigma$ and grows linearly with time for any $\sigma$,
ie. $\mathcal{A}_t\simeq 2\pi t$, so that the existence of a limiting
set $\hat \mathbb{K}_t$ is not ruled out.

It is implicit in the previous conjecture that the uniformizing maps
$f_t$ have a finite limit, say $\hat f_t$, as $\sigma\to 0$ such that
the image of the unit circle by $\hat f_t$ is the boundary of
$\mathbb{C}\setminus\hat \mathbb{K}_t$.  This leads us to conjecture
that the limiting domain $\hat \mathbb{O}_t$, which is the component
of $\mathbb{C}\setminus \hat \mathbb{K}_t$ containing infinity, is
uniformized onto $\mathbb{D}$ by $\hat g_t\equiv \hat f_t^{-1}$.

Let $R_t$ be the conformal radius computed at finite $\sigma$ using
the maps $f_t$. Recall that if $D_{\rm hs}$ is the fractal dimension
we have $R_t^{D_{\rm hs}}\asymp \mathcal{A}_t$ for large $t$. By
dimensional analysis $R_t\simeq
\sigma(\mathcal{A}_t/\sigma^2)^{1/D_{\rm hs}}$ since $\sigma$ has
dimension of a {\tt [length]}.  Because in the HS problem the area
grows linearly with time without any dependence on $\sigma$, this also
reads:
$$ 
R_t^{D_{\rm hs}} \simeq \sigma^{D_{\rm hs}-2}\, t
$$
Since we expect the limiting domains and the limiting conformal
maps to exit,  $R_t$ should have a finite limit as $\sigma\to 0$ which 
coincides with the conformal radius $\hat R_t$ computed using $\hat f_t$.
As consequence, we expect:
$$
D_{\rm hs}=2
$$ 
That is: the fractal dimension of Laplacian growth clusters, 
within the hydrodynamic regularization, is $2$. If these conjectures
are true, they imply that the Hele-Shaw problem is not in the same
universality class as DLA.

These conjectures are compatible with that of ref.\cite{gener} based
on numerical studies of generalized iterated conformal maps. They are
not expected to apply to arbitrary regularizations because in
these cases the area may depend on the UV cutoff and the previous
dimensional analysis does not apply. It would then be
natural to wonder whether we may define renormalized uniformizing
maps in the limit of a vanishing cutoff.

\vskip 1.5 truecm

%% optional

\begin{acknowledgments}
  
  It is a pleasure thank Vincent Pasquier and Paul Wiegmann for
  discussions on this problem and the organizers of the Carg\`ese
  2004 ASI for providing us the opportunity to write these notes.

  Work supported in part by EC contract number HPRN-CT-2002-00325 of
  the EUCLID research training network.

\end{acknowledgments}

%% appendix optional
%  \chapappendix{This is the Appendix Title}
%  This is an appendix with a title.
%  \chapappendix{}
%  This is an appendix without a title.

\end{document}